\documentclass[12pt]{article}

\usepackage[draft,showcomments]{ajtex}
\usepackage{physics}

\usepackage{erewhon,inputenc}
\usepackage[hang,flushmargin]{footmisc} 

\newcommand{\be}{\begin{equation}}
\newcommand{\ee}{\end{equation}}

\newcommand{\bit}{\begin{itemize}}  \newcommand{\eit}{\end{itemize}}

\let\latexd\d

\def\one{\mbox{1 \kern-.59em {\rm l}}}

\definecolor{green}{rgb}{0.0, 0.5, 0.0}
\definecolor{violet}{rgb}{0.58, 0.0, 0.83}
\definecolor{magenta}{rgb}{0.78, 0.08, 0.52}
\definecolor{dodgerblue}{rgb}{0.12, 0.56, 1.0}

\definecolor{beaver}{rgb}{0.62, 0.51, 0.44}
\definecolor{airforceblue}{rgb}{0.36, 0.54, 0.66}

\def\d{\delta}    
\def\e{\epsilon}

\def\i{\iota}

\def\l{{\lambda}} 
\def\m{\mu} \def\n{\nu}

\def\lb{{\overline l}}

\def\d{\delta}

\def\pa{\partial} 

\def\uno{\mbox{1 \kern-.59em {\rm l}}}

\newcommand{\0}{\,\!}      %this is just NOTHING!
\def\one{1\!\!1\,\,}

\def\Box{\square}

\def\bcomment#1{}
\def\IR{\relax{\rm I\kern-.18em R}}

\def \3{{(3)}}

\def \3{{(3)}}
\def \6{{(6)}}
\def \2{{(2)}}
\def \7{{(7)}}
\def \4{{(4)}}
\def\1{{(1)}}
\def\5{{(5)}}
\def\0{{(0)}}

\def\det{\text{det}}
\def\sgn{\text{sgn}}

\def\w{{\wedge}}

\def\rb{{\bar{r}}}

\renewcommand{\lb}{\left(}
\renewcommand{\rb}{\right)}

\title{Manifestly Covariant Polynomial M5-brane Lagrangians}

\author{Suk\latexd{r}ti Bansal}\email{sukrti.b@gmail.com}

\affiliation{\small Institute for Theoretical Physics, Karlsruhe Institute of Technology, Wolfgang-Gaede-Stra{\ss}e 1, 76131 Karlsruhe, Germany}

\affiliation{\small Institute for Theoretical Physics, Heidelberg University, Philosophenweg 19, 69120 Heidelberg, Germany}

\abstract{We present polynomial and manifestly covariant M5-brane Lagrangians along with their analyses involving their dynamics, gauge symmetries and their nonlinear selfduality condition. Such Lagrangians can be particularly useful for developments that are otherwise hindered by a non-polynomial structure and singularity of the Lagrangian such as its quantisation. Although on integrating out some of the auxiliary fields these polynomial Lagrangians reduce to the M5-brane Lagrangian given by the Pasti-Sorokin-Tonin (PST) formalism, in the analysis of the polynomial Lagrangians the only remnant of the non-polynomial structure of the PST type Lagrangian appears in the gauge transformation corresponding to an infinitesimal shift of a St\"uckelberg field. This transformation does not affect the dynamics or the on-shell self-duality condition of the polynomial M5-brane Lagrangians.}

\begin{document}
% \font\cmss=cmss10 \font\cmsss=cmss10 at 7pt

\maketitle

%\thispagestyle{empty}

%\newpage

%\vfill
%\vskip 5.mm
%\hrule width 5.cm
%\vskip 2.mm
%\scriptsize
% \newpage

\section{Introduction}

% \vspace{-0.35em}

The solutions of the low-energy limit of M-theory, i.e. 11D supergravity, consist of two branes -- the M2-brane and the M5-brane. An M2-brane is electrically charged under the supergravity field $C_3$ while its electromagnetic dual, an M5-brane, is charged magnetically. In the absence of compactification the only known supersymmetric excitations of M-theory are the M2-brane, the M5-brane and three fields -- a $3$-form gauge field $C_3$, a graviton and a gravitino. An M5-brane carries a self-dual (or chiral) $2$-form gauge field on its worldvolume.

%describing a stack of M2-branes is the ABJM superconformal field theory \cite{Aharony:2008ug} which is well-understood by now. 
The first description of the worldvolume theory of two coincident M2-branes was given by the maximally supersymmetric BLG model \cite{Bagger:2007vi,Bagger:2007jr,Bagger:2006sk,Gustavsson:2007vu,Gustavsson:2008dy}. Soon after, it was generalised to an arbitrary number of M2-branes by the ABJM superconformal field theory \cite{Aharony:2008ug}. The $3$D worldvolume theory of a stack of M2-branes is well-understood by now. However, the $6$D worldvolume theory of a stack of M5-branes, known as the (2,0) superconformal field theory, is yet to be fully comprehended. Its non-abelian symmetry structure and nonlinear self-duality conditions obscure its understanding. For many reasons (see for e.g. \cite{Lambert:2019diy}) it is possible that a classical Lagrangian for this theory simply does not exist and this theory can be described only at the quantum level. Despite its evasiveness the (2,0) superconformal field theory is worth pursuing as it enables the deduction of a large number of properties and dualities of lower-dimensional effective quantum field theories.

%(2,0) signifies the presence of two supercharges of the same chirality in the theory. 
% Later when I talk about interactions: The worldvolume field theories for coincident M2-
% and M5-branes are highly nontrivial, as the symmetry group is expected to be nonabelianized if there are multiple branes on top of each other. Moreover, the worldvolume theories are expected to be strongly interacting, as M2- and M5-branes can
% be viewed as the strong coupling limits of D2- and D4-branes through IIA M-theory
% duality. There are already non-abelian theories for multiple M2-branes [48{52] that
% can produce the N3=2 entropy behaviour [53]. However, it is fair to say that the
% non-abelian structure for multiple M5 branes is still an open question, though there
% are quite a few models on the market [54{64]. M-theory postulates the existence of strongly coupled 3d
% and 6d maximally supersymmetric conformal field theories [69{71], which are the
% worldvolume field theories of M2- and M5-branes respectively in the decoupling limit
% of gravity. 

% Nevertheless, even in the abelian case which corresponds to a single M5-brane, the complete formulation of a supersymmetric self-interacting worldvolume theory is already nontrivial. The field equations were first derived by the super-embedding formulation [83, 84]. It was later realised that they can also be derived from action principles [13,14]. It was then found that these descriptions of the M5-brane are all equivalent [85, 86].

In the context of developing and studying the action of a single M5-brane, initially the super-embedding formulation was used to determine the field equations \cite{Howe:1996yn,Howe:1997fb} which were later derived using the action principle \cite{Pasti:1997gx,Bandos:1997ui,Aganagic:1997zq}. These different formulations were shown to be equivalent in \cite{Bandos:1997gm,Howe:1997vn}. M5-brane models based on the Nambu-Poisson bracket are given in \cite{DeCastro:2003xop,Ho:2008ve}. Although it is easier to approach a model of a single M5-brane than that of multiple M5-branes, construction of the Lagrangian of a single M5-brane has had its own share of problems -- lack of manifest invariance under diffeomorphism \cite{Aganagic:1997zq,Ko:2016cpw,Schwarz:1997mc}, being non-polynomial \cite{Pasti:1997gx,Bandos:1997ui,Ko:2013dka}, having an infinite number of auxiliary fields \cite{Berkovits:1996rt,Bengtsson:1996fm,Devecchi:1996cp}, requiring going to a higher dimension \cite{Witten:1996hc,Witten:1999vg}, etc. These are more generally also the issues associated with most theories invariant under electromagnetic duality. Such hurdles impede further developments such as quantising the action and understanding its interactions.

In recent years there has been considerable interest in the study of M5-brane actions (e.g. \cite{Lambert:2010iw,Lambert:2011gb,Fiorenza:2012tb,Ko:2015zsy,Ko:2016cpw,Ko:2017tgo,Samann:2017sxo,Samann:2019eei,Rist:2020uaa,Bak:2019lye,Lambert:2019khh,Lambert:2019jwi,Lambert:2020scy,Andriolo:2020ykk,Vanichchapongjaroen:2020wza,Phonchantuek:2023iao,Gustavsson:2019ick,Gustavsson:2019efu,Gustavsson:2020ugb,Gustavsson:2021iex,Gustavsson:2022jpo,Gustavsson:2023zny,Andrianopoli:2022bzr}). In light of string field theory, Ashoke Sen put forward a new formalism for duality-symmetric actions in \cite{Sen:2015nph,Sen:2019qit} which describes abelian self-dual $2n$-form fields in $(2n+2)$ spacetime dimensions. The shortcomings of the prior approaches to construct duality-symmetric and M5-brane actions are avoided by Sen's formalism. It realises self-duality off-shell and preserves Lorentz invariance by introducing an unphysical auxiliary field which decouples from the dynamics. Initially this formalism was shown on a flat background in the weak gravity limit but recently it was extended to a general background \cite{Hull:2023dgp}.

% Sen's formalism has been used for many further developments including the formulation of M5-brane actions
% \cite{Andriolo:2020ykk} explored different geometrical aspects of an abelian (2,0) superconformal lagrangian with a chiral 2-form field while including matter. Their outcomes are as anticipated for the low-energy physics of an M5-brane action
% {\color{red} write about connection with M5-branes here}
Among other developments, Sen's formalism has been employed in the direction of constructing M5-brane actions \cite{Lambert:2019diy,Andriolo:2020ykk,Vanichchapongjaroen:2020wza,Phonchantuek:2023iao,Rist:2020uaa,Gustavsson:2020ugb,Andrianopoli:2022bzr}. In \cite{Lambert:2019diy} Sen's model was supersymmetrised to obtain a (2,0) action in Minkowski spacetime and showed its generalisation to an interacting non-abelian theory. It was further investigated in \cite{Andriolo:2020ykk} which explored different geometrical aspects of the abelian (2,0) superconformal action. The results turned out to be consistent with the low-energy physics of an M5-brane action. In \cite{Vanichchapongjaroen:2020wza} Sen's formalism was extended to construct a complete M5-brane action using the Green-Schwarz formalism, where the only manifest supersymmetry exists in the target space.
In one of the recent M5-brane models \cite{Rist:2020uaa} the mathematical origin of Sen's action was provided by employing the language of homotopy Maurer–Cartan theory. \cite{Andrianopoli:2022bzr} presented a rheonomic Lagrangian describing non-interacting tensor
multiplets on coincident M5-branes in superspace. As exhibited by all of these works, Sen's formalism has proved to be very useful.

While Sen's formalism realises Lorentz invariance manifestly, the same is not true of diffeomorphism invariance, which is realised in an unconventional and involved way. In \cite{Hull:2023dgp} Hull put forth an action that is a generalisation of Sen's duality-symmetric action \cite{Sen:2015nph,Sen:2019qit}. Hull's theory can be formulated geometrically on any manifold and is invariant under standard diffeomorphisms. The non-standard diffeomorpshim in Sen's action turns out to be a consequence of restricting the auxiliary sector metric in Hull's action to be the Minkowski metric.

The PST formulation of a duality invariant action \cite{Pasti:1995ii,Pasti:1995tn,Pasti:1996vs}, which is an alternative to Sen's formulation, has most of the characteristics required of a duality-invariant action like manifest covariance, finite number of auxiliary fields, etc. but it is not polynomial. It has derivatives of a scalar field present in its denominator making the Lagrangian singular when the scalar field is constant in spacetime. In \cite{Mkrtchyan:2019opf,Bansal:2021bis}, a polynomial formulation of the PST Lagrangian\footnote{not to be confused with the PST M5-brane Lagrangian} was shown.

In this paper we present a class of M5-brane Lagrangians which circumvent all of the earlier mentioned problems associated with M5-brane Lagrangians, and serve as an alternative to the M5-brane Lagrangians based on Sen's formulation. The Lagrangians given here are manifestly covariant, polynomial and have a finite number of auxiliary fields. Although these Lagrangians can be shown to reduce to the PST type non-polynomial M5-brane Lagrangian \cite{Pasti:1997gx,Bandos:1997ui} by solving for and eliminating specific auxiliary fields, these polynomial Lagrangians remain regular throughout, unlike the earlier non-polynomial Lagrangian. We find that the only remnant of the non-polynomial structure of the earlier Lagrangian \cite{Pasti:1997gx,Bandos:1997ui} occurring in our model of polynomial Lagrangians, is in the gauge symmetry linked with an infinitesimal shift of the St\"uckelberg field $a$. But this does not affect the dynamics or the nonlinear self-duality condition of the polynomial Lagrangians.

%which was introduced to make the Lagrangian diffeomorphism invariant

The polynomial Lagrangians are analytic and devoid of issues like discontinuity and singularity. They can be useful for the construction of the Hamiltonian formulation of an M5-brane \cite{Bergshoeff:1998vx}, quantisation \cite{Martin:1994np,Dijkgraaf:1996hk,Andriolo:2021gen} and extension to the non-abelian case \cite{Gustavsson:2020ugb,Bak:2019lye}.  They can also be more amenable in performing the dimensional reduction of M5-brane actions to D-brane actions \cite{Berman:1998va,Nurmagambetov:1998gp,Gustavsson:2021iex,Phonchantuek:2023iao} and in the search for the interactions of M5-branes \cite{Perry:1996mk}. 

The outline of this paper is as follows. In Sec. \ref{J12} we introduce a polynomial M5-brane Lagrangian in which an auxiliary $2$-form field appears via its field strength denoted by $G$. We discuss the dynamics, symmetries and the self-duality condition of that Lagrangian. In Sec. \ref{H12} we show the analysis of the polynomial M5-brane Lagrangian expressed in terms of the gauge potentials of the auxiliary $2$-form fields through a field redefinition. Sec. \ref{Lag_nonpol} shows the non-polynomial M5-brane Lagrangian obtained on eliminating the auxiliary $2$-form fields from the Lagrangian in Sec. \ref{H12}. Then in Sec. \ref{earl} we demonstrate the equivalence of the Lagrangians given in Secs. \ref{J12} - \ref{Lag_nonpol} with the PST M5-brane Lagrangian \cite{Pasti:1997gx,Bandos:1997ui}. Our results and possible future directions are discussed in Sec. \ref{disc}. In Appendix \ref{alt_poly} we show a few other polynomial M5-brane Lagrangians alternative to the Lagrangians presented in Secs. \ref{J12} and \ref{H12}. Appendix \ref{convDF} summarises our conventions and notation. Appendices \ref{app:attemptEa} and \ref{determinant} contain the proofs for certain identities used in obtaining some of the results in this paper.

\section{A Polynomial M5-brane Lagrangian with Field $J$}\label{J12}

We begin with introducing the background spacetimes. The worldvolume of an M5-brane is six-dimensional with the local coordinates denoted by $x^\mu\, (\mu=0,1,...5)$. The coordinates of $11$D supergravity background are given by $Z^M=(X^m,\theta^\alpha)$ where $X^m$ are the bosonic coordinates with $m=0,1,...10$ and $\theta^\alpha$ are the Majorana spinor coordinates with $\alpha=1,2,...32$. The supervielbein we use to switch between the curved spacetime and the tangent spacetime is $E_M{}^A(Z)$. Here $M$ denotes the curved spacetime indices and $A$ the tangent spacetime indices. The $11$D metric is the Minkowski metric $\eta_{MN}(Z)$. The $6$D worldvolume metric of the M5-brane is written as $g_{\m\n}(x)=E_\m{}^A E_\n{}^B\eta_{AB}$. Both $g_{\m\n}(x)$ and $\eta_{MN}(Z)$ have the mostly-plus signature.

Now we look at the field content. The chiral gauge potential of the M5-brane is denoted by the $2$-form field $A(x)$ and its field strength $F(x)=dA(x)$. The M5-brane is coupled to the $3$-form field $C_3(Z)$ and the $6$-form field $C_6(Z)$, which are the background fields in $11$D supergravity. The field strength of $C_6$ is dual to that of $C_3$ as follows:
\begin{align}
    dC_6={*}dC_3-\tfrac12\,C_3\w dC_3\,.
\end{align}
Apart from the three fields $A(x), C_3(Z)$ and $C_6(Z)$ we have five auxiliary fields $-$ three scalar fields $a(x),b(x),c(x)$ and two $2$-form fields $B(x)$ and $K(x)$. The field strength of $B(x)$ is $G(x)=dB(x)$. A manifestly covariant M5-brane Lagrangian which is polynomial, is given by
% \begin{align}\label{Lag_J12}
%     \mathcal{L}_{J12}
%     = &\,\,K\w[{*}\i_{\pa a}(J-2\,c\,{*}J)+\tfrac{c}{2}K\w{*}(\i_{\pa a}{*J}\w\i_{\pa a}{*J})] +H\w(J-C_3)-C_6 +2\,(c-c\,b^2-b){*1}
% \end{align}
\begin{align}\label{Lag_J12}
    \mathcal{L}_{J12} = da\w K\w[{*}J-2\,c\, J+\tfrac12\,c\,K\w{*}({*}J\w\i_{\pa a}{*}J)] +H \w (J-C_3)-C_6 +2\,{*}(c-c\,b^2-b).
\end{align}
% \begin{align}
%     \mathcal{L}_{J12}
%     =da \w  K \w[{*}J -2\,c\, J -\tfrac12\,c\,K \w {*}\{{*}J \w {*}(da \w  J)\}]  + H \w (J-C_3) -C_6  + 2\,(c - c\,b^2 - b)\,{*1}
% \end{align}
Here $H=(F+C_3)$ is referred to as the generalised field-strength of $A$. The field $J=(H+a\,G)$ is introduced for conciseness of notation. In the above Lagrangian $\i_{\pa a}{*J}$ is the interior product of ${*J}$ with the vector field $\pa a$\footnote{This notation is in accordance with the standard notation wherein a vector field $X$ is given by $X=X^\m\,\pa_\m\,$.}$\,=\pa^\mu a\,\pa_\mu\,$. $*$ denotes the Hodge dual (See Appendix \ref{sec:forms-appendix}). By identity \eqref{ivstarid} $\,\i_{\pa a}{*J}$ can also be written as $\,-\,{*}(da\w J)$ where $da=\pa_\m a\,dx^\m=\pa a^\flat$ is a $1$-form (For details see Appendix \ref{ext_algeb}). The term $(H\w C_3+C_6)$ in Lagrangian \eqref{Lag_J12} is the Wess-Zumino term.

%with the vector field $\pa^\mu a\,\pa_\mu$ which is denoted by $\pa a$
% Please note that in our notation $\pa a\neq \pa^\m a$ or $\pa_\m a\,$.

The two terms in Lagrangian \eqref{Lag_J12} which contain the $2$-form $K$ as well as the scalar $c$, have the field $J$ appearing in them once and twice respectively. Therefore the Lagrangian is labelled with the subscript $\,\text{`}12\,\text{'}\,$. By changing the number of the $J$ fields in those two terms of the Lagrangian it is possible to get a total of eight different Lagrangians: ${\cal L}_{J00},$ ${\cal L}_{J01},$ ${\cal L}_{J02},$ ${\cal L}_{J03},$ ${\cal L}_{J10},$ ${\cal L}_{J11},$ ${\cal L}_{J12},$ ${\cal L}_{J13}.$ Four of them are shown in Appendix \ref{alt_poly}. The subscript nomenclature is useful to distinguish between such Lagrangians.

% The subscript $\,\text{`}12\,\text{'}\,$ in $\,{\cal L}_{J12}\,$ denotes the number of $J$ fields in the second quadratic term and the quartic term of $\,{\cal L}_{J12}\,$ respectively. As we can see in Appendix \ref{alt_poly} there are a couple of different ways of writing a polynomial M5-brane Lagrangian in this formalism. They mainly differ by the number of the $J\,$ (or $H$)$\,$ fields appearing in their second quadratic and quartic terms. The subscript nomenclature is useful to distinguish between such Lagrangians.

% As we will see in Sec. [], eliminating the scalar fields $b$ and $c$ by plugging their solutions into the above Lagrangian gives us the Dirac-Born-Infeld (DBI) term which has a square-root.

The M5-brane action is expressed as
\begin{align}
    S_{J12}=\int_{{\cal M}_6}\!\! {\cal L}_{J12}\,,
\end{align}
where ${\cal M}_6$ is the $6$D worldvolume manifold of the M5-brane.

The equations of motion of Lagrangian \eqref{Lag_J12} are as follows:
\allowdisplaybreaks
\begin{align}\label{eom_J12}
    &E_a\equiv d[K\w\{{*}J-2\,c\,J+c\,K\w{*}({*}J\w\i_{\pa a}{*J})\}]+G\w(J-\i_{\pa a}{*K}-2\,c\,da\w K) \nn\\
    &\qquad\quad-c\,da\w K\w K\w{*}(\i_{\pa a}{*G}\w\i_{\pa a}{*J})=0\,, \nn\\
    &E_b\equiv 2\,b\,c+1=0 \,,\nn\\
    &E_c\equiv K\w[\,2\,da\w J- \tfrac{1}{2}\, K\w{*}(\i_{\pa a}{*}J\w\i_{\pa a}{*}J)]+2\,{*}(b^2-1)=0\,, \nn\\
    &E_A\equiv d\big[\i_{\pa a}{*}K-H+da\w[2\,c\, K+c \,{*}\{{*}(K\w K)\w\i_{\pa a}{*}J\}]+a\,G\big]=0\,, \nn\\
    &E_B\equiv d\big[a\,\{\i_{\pa a}{*}K-H+da\w[\,2\,c\, K+c\,{*}\{{*}(K\w K)\w\i_{\pa a}{*J}\}]\}\big]=0\,, \nn\\
    &E_K\equiv da\w[{*}J-2\,c\,J+c\,K\w{*}({*J}\w\i_{\pa a}{*J})]=0\,, \nn\\
    &E_{C_3}\equiv \i_{\pa a}{*} K+F+da\w[\,2\,c\, K+c \,{*}\{{*}(K\w K)\w\i_{\pa a}{*}J\}]+a\,G=0\,.
\end{align}
They describe the dynamics of Lagrangian ${\cal L}_{J12}$.

% \begin{align}
%     E_B\equiv\,& d\big[\,a\,\i_{\pa a}{*}K+da\w[2\,c\,a\, K+c\,a\,{*}\{{*}(K\w K)\w\i_{\pa a}{*J}\}]-aH\big]=0
% \end{align}

\subsection{Gauge Symmetries of ${\cal L}_{J12}$}\label{gauge_J12}

% They tell us about the va in transforming the fields without altering the Lagrangian. 
In this subsection we look at the gauge symmetries of Lagrangian ${\cal L}_{J12}$. As the fields $A$ and $B$ appear in \eqref{Lag_J12} only through their field strengths $F$ and $G$, the Lagrangian is invariant under the following abelian gauge transformations which are independent of each other:
% $\d A= dY_1$ and $\d B= dY_2\,$, with $Y_1$ and $Y_2$ being arbitrary $1$-forms.
\begin{align}
    \d_d A&= dY_1\,, \label{abelianA_J12} \\
    \d_d B&= dY_2\,, \label{abelianB_J12}
\end{align}
with $Y_1$ and $Y_2$ being arbitrary $1$-forms.

It can be seen that $J=(H+a\,G)=(dA+C_3+a\,dB)$, as well as Lagrangian ${\cal L}_{J12}$, are invariant under the gauge transformation
\begin{align}\label{AB_trans_J12}
    \d A=-\,a\,da\w Y_3\,,\quad \d B=da\w Y_3\,,\,
\end{align}
% Using identities \eqref{ivstarid} and \eqref{starivid}, Lagrangian \eqref{Lag_J12} can be rewritten as
% \begin{align}\label{Lag_J12'}
%     \mathcal{L}'_{J12}
%     = &\,\, K\w da\w[{*}J-2\,c\, J-\tfrac{c}{2}\,K\w{*}\{{*}J\w{*}(da\w J)\}] +H\w(J-C_3)-C_6 \nn\\
%     & +2\,(c-c\,b^2-b){*1}
% \end{align}
where $\,Y_3\,$ is an arbitrary $1$-form. Every term in ${\cal L}_{J12}$ with the field $K$ has an exterior product with $da\,$. We know that $da\w da=0$. Therefore, shifting the field $K\,$ by an exterior product with $da$:
\begin{align}
    \d_a K=da\w Y_4\,\, \text{where}\,\, Y_4\,\, \text{is an arbitrary}\, 1\text{-form}, \label{K_trans_J12}
\end{align}
does not alter the Lagrangian. The auxiliary field $a$ is a St\"uckelberg field as it makes the Lagrangian diffeomorphism invariant. When the field $a$ shifts by an infinitesimal scalar $\varphi$, the other fields transform in the following manner to give a gauge symmetry:
% \begin{align}
%     &\d_\varphi a=\varphi\,, \quad\d_\varphi b=\frac{(2\,b\,c-1)}{2\,\i_{\pa a}(da)}\,\i_{\pa a}\i_{\pa\varphi}{*}(K\w\i_{\pa a}{*}H)\,,\quad\d_\varphi c=-\,\frac{2\,c^2}{\i_{\pa a}(da)}\,\i_{\pa a}\i_{\pa\varphi}{*}(K\w\i_{\pa a}{*}H)\,, \nn\\
%     &\d_\varphi A=-\,a\,\d_\varphi B\,, \quad \d_\varphi K=-\,\i_{\pa a}\Big[{*}\Big(\frac{d\varphi\w\i_{\pa a}H}{[\i_{\pa a}(da)]^2}\Big)+d\varphi\w K\Big]\,,\nn\\
%     &\d_\varphi B=\frac{\i_{\pa a}}{\i_{\pa a}(da)}{*}\big[\,c\,(\d_{\varphi}K)\w{*}({*H}\w\i_{\pa a}{*H})-c\,K\w{*}\{{*}H\w{*}(d\varphi\w H)\}-{*}(d\varphi\w B)\nn\\
%     &\qquad\qquad\qquad-\d_\varphi c\,\{2\,H-K\w{*}({*}H\w\i_{\pa a}{*}H)\}\big]\,.
% \end{align}
\begin{align}\label{a_trans_J12}
    &\d_\varphi a=\varphi\,, \quad\d_\varphi b=\frac{(2\,b\,c-1)}{2\,\i_{\pa a}(da)}\,\i_{\pa a}\i_{\pa\varphi}{*}(K\w\i_{\pa a}{*}J)\,,\quad\d_\varphi c=-\,\frac{2\,c^2}{\i_{\pa a}(da)}\,\i_{\pa a}\i_{\pa\varphi}{*}(K\w\i_{\pa a}{*}J)\,, \nn\\
    &\d_\varphi A=-\,a\,\d_\varphi B\,, \quad \d_\varphi K=-\,\frac{\i_{\pa a}}{\i_{\pa a}(da)}\Big[\frac{\i_{\pa \varphi}{*}(\i_{\pa a}J)}{\i_{\pa a}(da)}-\varphi\,{*}G+d\varphi\w K\Big]\,,\nn\\
    &\d_\varphi B=-\,\frac{\i_{\pa a}}{\i_{\pa a}(da)}{*}\Big[\frac{d\varphi\w\i_{\pa a}\{{*}J-2\,c\,J+c\,K\w{*}({*}J\w\i_{\pa a}{*}J)\}}{\i_{\pa a}(da)}+{*}(2\,d\varphi\w B+\varphi\,dB)\nn\\
    &\qquad\qquad\qquad\qquad+c\,d(\varphi B)-2\,c\,K\w{*}\{{*}d(\varphi B)\w\i_{\pa a}{*J}\}-c\,K\w{*}({*}J\w\i_{\pa a}{*}J)\nn\\
    &\qquad\qquad\qquad\qquad-c\,(\d_{\varphi}K)\w{*}({*J}\w\i_{\pa a}{*J})-\d_\varphi c\,\{2\,J-K\w{*}({*}J\w\i_{\pa a}{*}J)\}\Big]\,.
\end{align}
% While we can see that when $\,\i_{\pa a}(da)=\pa_\m a\,\pa^\m a=0\,$ in the above gauge transformation the transformations of all the fields except $a$, become singular, it is important to note that $\,\pa_\m a\,\pa^\m a=0\,$ also implies $\d_\varphi a=0$. When $\d_\varphi a=0$ there is not shift in $a$ and the above transformation simply does not take place. Therefore transformation \eqref{a_trans_J12} is non-singular in its domain. Hence the Lagrangian system remains analytic throughout.
Due to the presence of $\,\i_{\pa a}(da)=\pa^\m a\,\pa_\m a\,$ in the denominator, the above gauge symmetry does not hold when $\,\i_{\pa a}(da)=0$, which happens when the scalar field $a(x)$ is constant in spacetime. Transformation \eqref{a_trans_J12} is the only non-polynomial gauge transformation of Lagrangian ${\cal L}_{J12}$. The general form of the solution to polynomial Lagrangian \eqref{Lag_J12} is represented by its nonlinear self-duality relation \eqref{selfdual_J12}. As we will see in Sec. \ref{nselfdual_J12}, derivation of the nonlinear self-duality relation of the M5-brane Lagrangian makes use of gauge transformations \eqref{abelianB_J12} and \eqref{AB_trans_J12}. Hence transformation \eqref{a_trans_J12} does not have any bearing on the general solution or the polynomial nature of Lagrangian ${\cal L}_{J12}$. This transformation simply depicts the gauge freedom of the St\"uckelberg scalar field $a$.

%, which does not affect the properties of the physical fields.

%The presence of $\,\i_{\pa a}(da)=\pa_\m a\,\pa^\m a\,$ in the denominator makes the above gauge transformation non-polynomial. 

The background gauge fields $C_3$ and $C_6$ can also be transformed in a way such that Lagrangian \eqref{Lag_J12} remains invariant:
\begin{align}\label{C_trans_J12}
    \d_W A=-W\,,\quad \d_W C_3 =dW\,,\quad \d_W C_6 =dW\w C_3\,\, \text{where}\,\, W\,\, \text{is an arbitrary}\, 2\text{-form}.
\end{align}

All of these gauge transformations form a closed algebra.

\subsection{Nonlinear Self-Duality Condition for ${\cal L}_{J12}$}\label{nselfdual_J12}

%So we have the freedom to apply the gauge transformations to those equations of motion, any of their combinations and the Lagrangian.

%We can arrive at the general form of the solution to Lagrangian ${\cal L}_{J12}\,$ by using those equations of motion in \eqref{eom_J12} and some of the gauge-symmetries in Sec. \ref{gauge_J12}.

Except the equation of motion of the St\"uckelberg field $a$ which lacks independent dynamics, the equations of motion and the Lagrangian are invariant under the gauge transformations given in Sec. \ref{gauge_J12}. So we have the freedom to apply the gauge transformations to the Lagrangian, the equations of motion of the fields other than $a$, and any combination of those equations of motion. Doing so we can arrive at the general form of the solution to Lagrangian ${\cal L}_{J12}\,$. As the gauge potential $A$ is a chiral field, it obeys a nonlinear self-duality condition, which gives the general condition satisfied by the solution.

Here we show the derivation of the nonlinear self-duality condition using the equations of motion of the fields $A$ and $B\,$ but other equations of motion barring $E_a$ can also be used if desired. For convenience we repeat the expressions of $E_A$ and $E_B$ below.
\begin{align}
    &E_A\equiv d\big[\i_{\pa a}{*}K-H+da\w[2\,c\, K+c \,{*}\{{*}(K\w K)\w\i_{\pa a}{*}J\}]+a\,G\big]=0\,, \nn\\
    &E_B\equiv d\big[a\,\{\i_{\pa a}{*}K-H+da\w[\,2\,c\, K+c\,{*}\{{*}(K\w K)\w\i_{\pa a}{*J}\}]\}\big]=0\,.
\end{align}
We consider a particular combination of these two equations, viz. $E_A-d[\i_{\pa a}(E_B-a\,E_A)/\i_{\pa a}(da)]=0,$ which gives
% We have,
% \begin{align}
%     E_B-a\,E_A=da\w(\i_{\pa a}{*}K-J)=0\,.
% \end{align}
% Taking the interior product of the above equation with $\pa a\,$ gives
% \begin{align}\label{JK_eq}
%    \i_{\pa a}{*}K-H=a\,G-\,\frac{da\w\i_{\pa a}J}{\i_{\pa a}(da)}.
% \end{align}
% Substituting the expression for $\,(\i_{\pa a}{*}K-H)$ from eq. \eqref{JK_eq} into $E_A$, we get,
\begin{align}\label{da_wedge_eqJ12}
    da\w d\bigg[\frac{\i_{\pa a}J}{\i_{\pa a}(da)}+2\,B-2\,c\,K-c\,{*}\{{*}(K\w K)\w\i_{\pa a}{*}J\}\bigg]=0.
\end{align}
The general solution to the equation $da\w dX=0\,$ where $X$ is a $\,p$-form, is $\,X=dY+da\w Z\,$ with $Y$ and $Z$ being arbitrary $(p-1)$-forms (see \cite{Bansal:2021bis}). So eq. \eqref{da_wedge_eqJ12} can be written as
\begin{align}
   \frac{\i_{\pa a}J}{\i_{\pa a}(da)}+2\,B-2\,c\,K-c\,{*}\{{*}(K\w K)\w\i_{\pa a}{*}J\}=dY+da\w Z,
\end{align}
where $Y$ and $Z$ are arbitrary $1$-forms. On performing gauge transformation \eqref{abelianB_J12} with $\d_d B=d(Y/2)\,$ and transformation \eqref{AB_trans_J12} such that $\,\d A=-\,a\,da\w (Z/2)\,,\, \d B=da\w (Z/2)\,$, the above equation gives us the nonlinear self-duality condition:
\begin{align}\label{selfdual_KBJ}
    \i_{\pa a}J=\{\i_{\pa a}(da)\}[\,2\,(c\,K-B)+c\,{*}\{{*}(K\w K)\w\i_{\pa a}{*}J\}]\,.
\end{align}
In view of gauge symmetry \eqref{K_trans_J12}, the term $da\w\i_{\pa a}K$ present in $\,{*}[da\w\i_{\pa a}(E_B-a\,E_A)]=0$, can be gauged away, effectively giving us
\begin{align}\label{Ksol_J12}
    K=\frac{\i_{\pa a}{*}J}{\i_{\pa a}(da)}\,.
\end{align}
% Absorbing this expression for $K$ into eq. \eqref{selfdual_KBJ} the nonlinear self-duality condition for ${\cal L}_{J12}$ gets expressed as:
% \begin{align}\label{selfdual_J12}
%     \i_{\pa a}J=\i_{\pa a}{*}\tilde{J}-2\,\{\i_{\pa a}(da)\}B,\quad \text{where}\,\,\tilde{J}=2\,c\,J-\frac{\,c\,\i_{\pa a}{*J}\w{*}({*J}\w\i_{\pa a}{*J})\,}{\i_{\pa a}(da)}\,.
% \end{align}
The field $B$ appears in Lagrangian ${\cal L}_{J12}$ and its equations of motion only via the field strength $G$ and not simply as the gauge potential $B$. So solving for $B$ is not straightforward here. However, it can be checked that $B$ can be expressed as:
\begin{align}\label{Bsol_J12}
    B=\frac{\i_{\pa a}J-2\,c\,\i_{\pa a}{*J}-c\,{*}[K\w{*}(\i_{\pa a}{*J}\w\i_{\pa a}{*J})]}{\i_{\pa a}(da)}\,.
\end{align}
This expression for $B$ is consistent with all the equations of motion. Absorbing the above two expressions for $K$ and $B$ into eq. \eqref{selfdual_KBJ}, the nonlinear self-duality condition for ${\cal L}_{J12}$ gets written as:
\begin{align}\label{selfdual_J12}
    \boxed{\i_{\pa a}J=\i_{\pa a}{*}\tilde{J}\,\, \Leftrightarrow\,\, da\w({*}J-\tilde J)=0}\quad \text{where}\,\,\tilde{J}=2\,c\,J-\frac{\,c\,\i_{\pa a}{*J}\w{*}({*J}\w\i_{\pa a}{*J})\,}{\i_{\pa a}(da)}\,.
\end{align}
This condition portrays the nonlinear chirality of the $2$-form gauge potential constituting the M5-brane. The anti self-duality condition $\i_{\pa a}J=-\,\i_{\pa a}{*}\tilde{J}$ can be obtained by redefining $\,b\,$ and $\,c\,$ in ${\cal L}_{J12}$ as $(-\,b)$ and $(-\,c)\,$ respectively. By virtue of the self-duality condition \eqref{selfdual_J12} it can be seen that the auxiliary field $B$ \eqref{Bsol_J12} vanishes on-shell. This implies that in $J=H+a\,dB$ we can set $dB=0\,$ which reduces self-duality condition \eqref{selfdual_J12} to
\begin{align}\label{selfdual_H12}
    \boxed{\i_{\pa a}H=\i_{\pa a}{*}\tilde{H}\,\, \Leftrightarrow\,\, da\w({*}H-\tilde H)=0}\quad \text{where}\,\,\tilde{H}=2\,c\,H-\frac{\,c\,\i_{\pa a}{*H}\w{*}({*H}\w\i_{\pa a}{*H})\,}{\i_{\pa a}(da)}.
\end{align}
Hence, we have seen the nonlinear duality symmetric structure of the polynomial Lagrangian ${\cal L}_{J12}$ describing an M5-brane. In the next section we introduce Lagrangian ${\cal L}_{H12}$, for which it is relatively easier to solve the equations of motion including the equation of motion of the field $B$. Consequently the simplification of the nonlinear self-duality condition is more tractable in the framework of ${\cal L}_H$ than of ${\cal L}_J$. We elaborate on this condition in Sec. \ref{nselfdual_H12}.

\section{A Polynomial M5-brane Lagrangian with Field $H$}\label{H12}

Now we look at an alternate formalism of Lagrangian ${\cal L}_{J12}$, in which the auxiliary field $B$ appears without an exterior derivative acting on it. For deriving this formalism we perform the field redefinition $A\rightarrow A-a\,B$ on Lagrangian ${\cal L}_{J12}$. It gives us
\begin{align}\label{Lag_H12}
    \mathcal{L}_{H12}
    =\,\,& da\w K\w[{*H}-2\,c\,H-\i_{\pa a}{*}B+\tfrac12\,c\,K\w{*}({*H}\w\i_{\pa a}{*}H)]-H\w(da\w B+C_3)-C_6\nn\\
    &+2\,{*}(c-c\,b^2-b)\,.
\end{align}
This Lagrangian appears with the generalised field strength $H$ whereas the expression of ${\cal L}_{J12}$ \eqref{Lag_J12} was in terms of the field $J$. Rewriting the above Lagrangian as below,
% \begin{align}
%     \mathcal{L}'_{H12}
%     =\,\,& da\w K\w[{*H}-2\,c\,H+\tfrac{c}{2}K\w{*}({*H}\w\i_{\pa a}{*}H)]+da\w B\w(H-\i_{\pa a}{*}K)-H\w C_3-C_6\nn\\
%     &-2\,b\,{*1}+2\,c\,(1-b^2)\,{*1}
% \end{align}
% \begin{align}
%     \mathcal{L}'_{H12}
%     =\,\,& da\w K\w{*H} -H\w C_3-C_6 +da\w B\w(H-\i_{\pa a}{*}K)-2\,b\,{*1}\nn\\
%     & +c\,[\,2-da\w K\w\{2\,H-\tfrac12\,K\w{*}({*H}\w\i_{\pa a}{*}H)\}-2\,b^2\,]\,{*1}\,,
% \end{align}
\begin{align}\label{Lag_H12'}
    \mathcal{L}_{H12}
    =\,\,& da\w K\w{*H} -H\w C_3-C_6 +da\w B\w(H-\i_{\pa a}{*}K)\nn\\
    & +c\,[\,da\w K\w\{ -\,2\,H+\tfrac{1}{2}\,K\w{*}({*}H\w\i_{\pa a}{*}H)\}+2\,{*}(1-b^2)\,]-2\,{*b}\,,
\end{align}
we can see that the auxiliary fields $B$ and $\,c\,$ are Lagrange multipliers. Solving the equation of motion of $B$ gives us the solution for $K$ and on solving the equation of motion $c$ we get the solutions for $b$.

% {\color{red} Solving their equations of motion (it seems it's not possible to see at this stage that $c$ is a Lagrange multiplier).}

% show ahead Solving the equation of motion of $B$ gives us the following solution for $K$ 
% \begin{align}\label{Ksol_H12}
%     K=\frac{\i_{\pa a}{*H}}{\i_{\pa a}(da)}\,.
% \end{align}
% In obtaining the above solution we have made use of the fact that terms of the kind $da\w Y^{(1)}$ do not contribute to the solution for $K$ due to the invariance of ${\cal L}_{H12}$ under the transformation $\d K=da\w Y^{(1)}$ (...). On plugging solution \eqref{Ksol_H12} into Lagrangian \eqref{Lag_H12} one can see that we get Lagrangian (...), a non-polynomial expression of the M5-brane Lagrangian.

The dynamics of ${\cal L}_{H12}$ is described by the following equations of motion:
\begin{align}\label{eom_H12}
    &E_a\equiv d[K\w({*H}-2\,c\,H-\i_{\pa a}{*}B) -c\,H\w{*}\{{*}(K\w K)\w\i_{\pa a}{*}H\}+B\w(H-\i_{\pa a}{*} K)]=0\,, \nn\\
    &E_b\equiv 2\,b\,c+1=0\,, \nn\\
    &E_c\equiv da\w K\w[\,2\, H- \tfrac{1}{2}\, K\w{*}({*}H\w\i_{\pa a}{*}H)]+2\,{*}(b^2-1)=0\,, \nn\\
    &E_A\equiv d\big[\i_{\pa a}{*}K-H+da\w[\,2\,c\, K+c\,{*}\{{*}(K\w K)\w\i_{\pa a}{*}H\}-B\,]\big]=0\,, \nn\\
    &E_B\equiv da\w(\i_{\pa a}{*K}-H)=0\,, \nn\\
    &E_K\equiv da\w[\,{*H}-2\,c\,H-\i_{\pa a}{*}B+c\,K\w{*}({*H}\w\i_{\pa a}{*}H)\,]=0\,, \nn\\
    &E_{C_3}\equiv \i_{\pa a}{*} K+F+da\w[\,2\,c\, K+c\,{*}\{{*}(K\w K)\w\i_{\pa a}{*} H\}-B\,]=0\,.
\end{align}
In comparison with the equations of motion of ${\cal L}_{J12}$ \eqref{eom_J12}, equations \eqref{eom_H12}, which are in terms of $H$ instead of the field $J$, are shorter and some of them are easier to solve, such as $E_a=0$ and $E_B=0$. 
% {\color{red}We elaborate in Sec.[] about solving the equations of motion.}

\subsection{Gauge Symmetries of ${\cal L}_{H12}$}

Here only the gauge potential $A$ has the abelian gauge symmetry wherein for an arbitrary $1$-form field $\,Y_1$, $\,{\cal L}_{H12}\,$ is invariant under
\begin{align}
    \d_d A= dY_1 \,. \label{abelian_H12}
\end{align}
If we transform gauge potential $A$ by $\,da\w Y_2\,$, with $Y_2$ being an arbitrary $1$-form, then $B$ can be transformed to compensate for $\,\d A\,$ in order to keep $\,{\cal L}_{H12}\,$ invariant as follows:
\begin{align}
    \d A=da\w Y_2\,,\quad \d B=-\,dY_2\,.\,  \label{AB_trans_H12}
\end{align}
In Lagrangian \eqref{Lag_H12'} we can see that all the terms with the fields $B$ and $K$ in them have an exterior product with $da\,$. Hence, using arbitrary $1$-forms $\,Y_3\,$ and $\,Y_4\,$, we have the following two independent gauge symmetries:
\begin{align}
    \d_a B&=da\w Y_3\,; \label{B_trans_H12}\\
    \d_a K&=da\w Y_4\,.  \label{K_trans_H12}
\end{align}
On giving an arbitrary infinitesimal shift to the St\"uckelberg field $a$ by another scalar field $\,\varphi\,$, we need to transform the other fields as shown below to keep the Lagrangian invariant:
% \begin{align}\label{a_trans_H12}
%     &\d_\varphi a=\varphi\,, \quad\d_\varphi b=\frac{(2\,b\,c-1)}{2\,\i_{\pa a}(da)}\,\i_{\pa a}\i_{\pa\varphi}{*}(K\w\i_{\pa a}{*}H)\,,\quad\d_\varphi c=-\,\frac{2\,c^2}{\i_{\pa a}(da)}\,\i_{\pa a}\i_{\pa\varphi}{*}(K\w\i_{\pa a}{*}H)\,, \nn\\
%     &\d_\varphi A=\varphi\,B\,, \quad \d_\varphi K=-\,\i_{\pa a}\Big[{*}\Big(\frac{d\varphi\w\i_{\pa a}H}{[\i_{\pa a}(da)]^2}\Big)+d\varphi\w K\Big]\,,\nn\\
%     &\d_\varphi B=\frac{\i_{\pa a}}{\i_{\pa a}(da)}{*}\big[\,c\,(\d_{\varphi}K)\w{*}({*H}\w\i_{\pa a}{*H})-c\,K\w{*}\{{*}H\w{*}(d\varphi\w H)\}-{*}(d\varphi\w B)\nn\\
%     &\qquad\qquad\qquad-\d_\varphi c\,\{2\,H-K\w{*}({*}H\w\i_{\pa a}{*}H)\}\big]\,.
% \end{align}
\begin{align}\label{a_trans_H12}
    &\d_\varphi a=\varphi\,, \quad\d_\varphi b=\frac{(2\,b\,c-1)}{2\,\i_{\pa a}(da)}\,\i_{\pa a}\i_{\pa\varphi}{*}(K\w\i_{\pa a}{*}H)\,,\quad\d_\varphi c=-\,\frac{2\,c^2}{\i_{\pa a}(da)}\,\i_{\pa a}\i_{\pa\varphi}{*}(K\w\i_{\pa a}{*}H)\,, \nn\\
    &\d_\varphi A=\varphi B\,, \quad \d_\varphi K=-\,\frac{\i_{\pa a}}{\i_{\pa a}(da)}\Big[\i_{\pa \varphi}{*}\Big\{\frac{\i_{\pa a}H}{\i_{\pa a}(da)}-B\Big\}-\varphi\,{*}(dB)+d\varphi\w K\Big]\,,\nn\\
    &\d_\varphi B=-\,\frac{\i_{\pa a}}{\i_{\pa a}(da)}{*}\Big[\frac{d\varphi\w\i_{\pa a}\{{*}H-2\,c\,H+c\,K\w{*}({*}H\w\i_{\pa a}{*}H)\}}{\i_{\pa a}(da)}+{*}(2\,d\varphi\w B+\varphi\,dB)\nn\\
    &\qquad\qquad\qquad\qquad+c\,d(\varphi B)-2\,c\,K\w{*}\{{*}d(\varphi B)\w\i_{\pa a}{*H}\}-c\,K\w{*}({*}H\w\i_{\pa a}{*}H)\nn\\
    &\qquad\qquad\qquad\qquad-c\,(\d_{\varphi}K)\w{*}({*H}\w\i_{\pa a}{*H})-\d_\varphi c\,\{2\,H-K\w{*}({*}H\w\i_{\pa a}{*}H)\}\Big]\,.
\end{align}
% This gauge transformation consists of the only set of non-polynomial expressions in the entire analysis of our polynomial M5-brane Lagrangian ${\cal L}_{H12}\,$. 
% here also this gauge transformation, stemming from a shift in $a$, can occur only when $\,\i_{\pa a}(da)=\pa_\m a\,\pa^\m a\neq 0\,$. Hence, despite its non-polynomial expression, transformation \eqref{a_trans_H12} is non-singular.

% If we transform all the fields by a particular gauge transformation while for every field $f$, $E_f=\d S/(\d f)$, then
% \begin{align}
%     \d a\, E_a+\d b\, E_b+\d c\, E_c+\d A\w E_A+\d B\w E_B+\d K\w E_K+\d C_3\w E_{C_3}+tot.\, der.=0.
% \end{align}
% where $tot.\, der.$ stands for the total derivative terms which are the boundary terms that vanish on integrating the Lagrangian to get the action. Considering gauge transformation \eqref{a_trans_H12} where $\,\d a=\varphi$, we get
% \begin{align}
%     E_a=-\,\frac{1}{\varphi}\,(\d_\varphi b\, E_b+\d_\varphi c\, E_c+\d_\varphi A\w E_A+\d_\varphi B\w E_B+\d_\varphi K\w E_K)+tot.\, der.
% \end{align}
% Thus gauge symmetry \eqref{a_trans_H12} helps us see that the dynamics of the St\"uckelberg field $a$ is determined by the dynamics of the other fields in the $6$D worldvolume of the M5-brane: $a$ is a gauge field with no independent dynamics of its own.
As in the case of ${\cal L}_{J12}$, here also this non-polynomial gauge symmetry stemming from an infinitesimal shift in the St\"uckelberg field $a$, does not affect the general form of the solution of ${\cal L}_{H12}\,$. If we transform every field $f$ by an infinitesimal gauge transformation $\d_g f$ while $E_f=\d {\cal L}/(\d f)${\footnote{neglecting the total derivative terms appearing in $(\d f\w E_f)$ as they vanish on integration and do not contribute to $\d S$}, then
% \begin{align}
%     \d a E_a+\d b E_b+\d c E_c+\d A\w E_A+\d B\w E_B+\d K\w E_K+\d C_3\w E_{C_3}=0.
% \end{align}
% Considering gauge transformation \eqref{a_trans_H12} where $\,\d a=\varphi\,$ with $\varphi$ being infinitesimally small, we get
% \begin{align}
%     E_a=-\,\frac{1}{\varphi}\,(\d_\varphi b\, E_b+\d_\varphi c\, E_c+\d_\varphi A\w E_A+\d_\varphi B\w E_B+\d_\varphi K\w E_K).
% \end{align}
\begin{align}
    \d_g S=\int_{{\cal M}_6}\!\!\d_g{\cal L}=\int_{{\cal M}_6}\!\!(\d_g a E_a{+}\d_g b E_b{+}\d_g c E_c{+}\d_g A\w E_A{+}\d_g B\w E_B{+}\d_g K\w E_K{+}\d_g C_3\w E_{C_3})=0.
\end{align}
Considering gauge transformation \eqref{a_trans_H12} where $\,\d_g a=\varphi\,$ with $\varphi$ being infinitesimally small, we get
\begin{align}
    \d_\varphi S=\int_{{\cal M}_6}\!\! \varphi\, \Big[E_a+\frac{1}{\varphi}\,(\d_\varphi b\, E_b+\d_\varphi c\, E_c+\d_\varphi A\w E_A+\d_\varphi B\w E_B+\d_\varphi K\w E_K)\Big]=0.
\end{align}
As $\varphi$ is an arbitrary infinitesimal parameter, by the fundamental lemma of calculus of variations the expression inside the square brackets is identically zero, i.e.,
\begin{align}
    E_a=-\,\frac{1}{\varphi}\,(\d_\varphi b\, E_b+\d_\varphi c\, E_c+\d_\varphi A\w E_A+\d_\varphi B\w E_B+\d_\varphi K\w E_K).
\end{align}
Thus gauge symmetry \eqref{a_trans_H12} helps us see that the dynamics of the St\"uckelberg field $a$ is determined by the dynamics of the other fields in the $6$D worldvolume of the M5-brane: $a$ is a gauge field with no independent dynamics of its own. $E_a=0$ is trivially satisfied by the solutions of the other equations of motion, and does not yield any new solution. It effectively acts as a conditional identity, the conditions being the equations of motion of all the fields except the St\"uckelberg field $a$.
% here also this non-polynomial gauge symmetry does not affect the general form of the solution or the polynomial form of ${\cal L}_{H12}\,$.

The gauge symmetry of $\,{\cal L}_{H12}\,$ involving the background gauge fields $C_3$ and $C_6$, is the same as the corresponding symmetry of $\,{\cal L}_{J12}\,$ \eqref{C_trans_J12}:
\begin{align}\label{C_trans_H12}
    \d_W A=-W\,,\quad \d_W C_3 =dW\,,\quad \d_W C_6 =dW\w C_3\,\, \text{where}\,\, W\,\, \text{is an arbitrary}\, 2\text{-form}.
\end{align}
% {\color{red} All of these gauge transformations form a closed algebra.}

\subsection{Nonlinear Self-Duality Condition for ${\cal L}_{H12}$}\label{nselfdual_H12}

As in Sec. \ref{nselfdual_J12} we again use $E_A$ and $E_B$ to obtain the general condition obeyed by the solution to ${\cal L}_{H12}$.
\begin{align}
    E_A\equiv\,&d\big[\i_{\pa a}{*}K-H+da\w[\,2\,c\, K+c\,{*}\{{*}(K\w K)\w\i_{\pa a}{*}H\}-B\,]\big]=0\,, \nn\\
    E_B\equiv\,& da\w(\i_{\pa a}{*K}-H)=0\,.
\end{align}
The equation $E_A-d[\i_{\pa a}E_B/\i_{\pa a}(da)]=0$ gives us
\begin{align}\label{E_B_in_E_A}
    &da\wedge d\Big[\frac{ \i_{\pa a}H}{\i_{\pa a}(da)}-2\,c\,K-c \,{*}\{{*}(K\w K)\w\i_{\pa a}{*}H\}+B\Big]=0.
\end{align}
As discussed in Sec. \ref{nselfdual_J12}, such an equation can be re-expressed as
\begin{align}
    &\frac{ \i_{\pa a}H}{\i_{\pa a}(da)}-2\,c\,K-c \,{*}\{{*}(K\w K)\w\i_{\pa a}{*}H\}+B=dX+da\wedge Y,
\end{align}
where $X$ and $Y$ are arbitrary $1$-forms. We apply gauge transformation \eqref{AB_trans_H12}$\,$ with $\,\d A=-\,da\w (X/2)\,$ and $\,\d B=d(X/2)\,$, getting
\begin{align}
    \frac{\i_{\pa a}H}{\i_{\pa a}(da)}-2\,c\,K-c\,{*}\{{*}(K\w K)\w\i_{\pa a}{*}H\}+B=da\w \Big[Y+\,\frac{\i_{\pa a}dX}{2\,\i_{\pa a}(da)}\Big].
\end{align}
Now if we transform the field $B$ in the above equation using \eqref{B_trans_H12} such that $\d_a B=da\w [Y+\,(\i_{\pa a}dX)/2\,\i_{\pa a}(da)]$, we get the following relation:
\begin{align}\label{1selfdual_H12}
    \i_{\pa a}H=\{\i_{\pa a}(da)\}[\,2\,c\,K-B+c\,{*}\{{*}(K\w K)\w\i_{\pa a}{*}H\}].
\end{align}
If one wishes to incorporate $E_K$ in the above derivation, the expression for $da\w B$ in the equation ${*}\i_{\pa a}E_K=0\,$ can be substituted into eq. \eqref{E_B_in_E_A}. Then gauge transformations \eqref{AB_trans_H12} and \eqref{B_trans_H12} can be applied along similar lines as above. Alternatively the solution for $B$ (which we get on solving $E_K=0$) can be plugged into relation \eqref{1selfdual_H12}. Using either of these two ways, we arrive at
\begin{align}\label{2selfdual_H12}
    \i_{\pa a}H=\{\i_{\pa a}(da)\}[\,c\,K+\tfrac12\,c\,{*}\{{*}(K\w K)\w\i_{\pa a}{*}H\}]+c\,\i_{\pa a}{*}H+\tfrac12\,c\,{*}\{K\w{*}(\i_{\pa a}{*H}\w\i_{\pa a}{*H})\}.
\end{align}
% plus terms not shown here as they vanish when substituted into the Lagrangian due to gauge symmetries \eqref{K_trans_H12} and \eqref{B_trans_H12}
Now we solve $E_B=0.$ It can be seen that ${*}E_B=0\,$ gives
\begin{align}\label{Ebsolv1}
    K=\frac{\i_{da}{*H}}{\i_{\pa a}(da)}+\frac{da\w\i_{da}K}{\i_{\pa a}(da)}.
\end{align}
$K=[\i_{da}{*H}/\i_{\pa a}(da)]$ is a particular solution of the above equation. Let the general solution be $K=[\i_{da}{*H}/\i_{\pa a}(da)]+Z$ where $Z$ is an as yet unknown $2$-form. Plugging this solution into eq. \eqref{Ebsolv1} gives $Z=[da\w\i_{\pa a}Z/\i_{\pa a}(da)].$ So the general solution can be written as
\begin{align}
    K=\frac{\i_{da}{*H}}{\i_{\pa a}(da)}+\frac{da\w\i_{da}Z}{\i_{\pa a}(da)}.
\end{align}
Applying gauge transformation \eqref{K_trans_H12} $\d_a K=da\w Y_4$ in the above equation, with the transformation parameter $Y_4$ gauge fixed to $\i_{\pa a}Z/\i_{\pa a}(da)$, we get the solution for $K$ as:
\begin{align}\label{Ksol_H12}
    K=\frac{\i_{da}{*H}}{\i_{\pa a}(da)}.
\end{align}
Similarly, $\,E_K=0\,$ can be solved to get:
\begin{align}
    B=\frac{\i_{\pa a}H-2\,c\,\i_{\pa a}{*H}-c\,{*}[K\w{*}(\i_{\pa a}{*H}\w\i_{\pa a}{*H})]}{\i_{\pa a}(da)}\,. \label{Bsol_H12}
\end{align}
Although we have shown the derivations of relation \eqref{1selfdual_H12} and the solution for $K$ \eqref{Ksol_H12} within the self-contained framework of ${\cal L}_{H12}$, these expressions can also be obtained by redefining $A\rightarrow A-a\,B$ in the corresponding expressions of ${\cal L}_{J12}$, viz. \eqref{selfdual_KBJ}-\eqref{Ksol_J12}. Inserting solutions \eqref{Ksol_H12}-\eqref{Bsol_H12} for $K$ and $B$ into \eqref{1selfdual_H12} gives the same relation as the one obtained on substituting the solution of $K$ into condition \eqref{2selfdual_H12}. The relation is the nonlinear self-duality condition in terms of the fields $A,C_3, a$ and $c\,$:
% \begin{align}
%     \i_{\pa a}H=\i_{\pa a}{*}\bigg[\,2\,c\,H-\frac{c\,\{\i_{\pa a}{*H}\w*({*H}\w\i_{\pa a}{*H})\}}{\i_{\pa a}(da)}\bigg].
% \end{align}
% \begin{align}
%     \i_{\pa a}(da)\,\i_{\pa a}H=\i_{\pa a}{*}[\,2\,c\,\i_{\pa a}(da)\,H-c\,\{\i_{\pa a}{*H}\w*({*H}\w\i_{\pa a}{*H})\}].
% \end{align}
\begin{align}\label{selfdual-cond}
    \boxed{\i_{\pa a}H=\i_{\pa a}{*}\tilde{H}\,\, \Leftrightarrow\,\, da\w({*}H-\tilde H)=0}\quad \text{where}\,\,\tilde{H}=2\,c\,H-\frac{\,c\,\i_{\pa a}{*H}\w{*}({*H}\w\i_{\pa a}{*H})\,}{\i_{\pa a}(da)}.
\end{align}
This self-duality relation is the same as that seen earlier in eq. \eqref{selfdual_H12}. This condition characterises the chiral nature of the field strength $H$ comprising the M5-brane. Here again, as in the case of ${\cal L}_{J12}$ in section \eqref{nselfdual_J12}, the solution for the auxiliary field $B$ \eqref{Bsol_H12} vanishes on-shell in view of the self-duality condition. If we consider Lagrangian \eqref{Lag_H12} with $\,b\rightarrow-b\,$ and $\,c\rightarrow-c\,$, then we get the anti self-duality condition $\i_{\pa a}H=-\,\i_{\pa a}{*}\tilde{H}$. Hence we have seen the gauge symmetries and the (anti) self-duality of the polynomial M5-brane Lagrangians.
% where $\tilde{H}=2\,c\,H-c\,\{\i_v{*H}\w*({*H}\w\i_v{*H})\}$.

%If we solve $E_B$ and $E_K$ and plug the solutions of $B$ and $K$ [] into eq. \eqref{selfdual_H12}, we get the nonlinear self-duality condition [] in terms of the fields $A, a$ and $c$.

\section{A Non-Polynomial M5-brane Lagrangian}\label{Lag_nonpol}

%which is devoid of the square-root present in the Dirac-Born-Infeld (DBI) term of a usual M5-brane Lagrangian. 

In this section we introduce a non-polynomial expression of the M5-brane Lagrangian. We get this Lagrangian by simply plugging solutions \eqref{Ksol_H12} - \eqref{Bsol_H12} for the fields $B$ and $K$ present in the previous section, into Lagrangian ${\cal L}_{H12}$ \eqref{Lag_H12}.
\begin{equation}\label{Lag_np}
    {\cal L}_{np}=\i_v{*H}\w[{*}(\i_vH)-2\,c\,{*}(\i_v{*H})+\tfrac12\,c\,\i_v{*H}\w{*}\big(\i_v{*H} \w\i_v{*H}\big)]-H\w C_3-C_6+2\,{*}(c-c\,b^2-b)
\end{equation}
% \begin{align}\label{Lag_np}
%     \mathcal{L}_{np}=\,\,&\i_v{*H}\w[{*}(\i_vH)-2\,c\,{*}(\i_v{*H})+\tfrac12\,c\,\i_v{*H}\w{*}\big(\i_v{*H}\w\i_v{*H}\big)]-H\w C_3-C_6\nn\\
%     &+2\,(c-c\,b^2-b)\,{*1}
% \end{align}
where $\,v=v^\mu\,\pa_\mu=(\pa^\m a/\sqrt{\pa^\n a\,\pa_\n a})\,\pa_\mu\,$ is a normalised vector-field.

% $v =da/\sqrt{\i_{\pa a}(da)}$
The dynamics of ${\cal L}_{np}$ can be seen via the following equations of motion:
% \begin{align}\label{M5eoms}
%     E_a &\equiv  d\bigg[\frac{v^\flat\w\i_v{*H}\w\i_v{*H}+v^\flat\w\i_vH\w[\i_v{H}-4\,c\,\i_v{*H}-2\,c\,{*}\{\i_v{*H}\w*(\i_v{*H}\w\i_v{*H})\}]}{\sqrt{\i_{\pa a}(da)}}\bigg]=0 \nn\\
%     E_b&\equiv 2\,b\,c+1=0 \nn\\
%     E_c&\equiv\i_v{*H}\w{*}(\i_v{*H})-\tfrac{1}{4}\,\i_v{*H}\w\i_v{*H}\w{*}(\i_v{*H}\w\i_v{*H})+(b^2-1){*1}=0 \nn\\
%     E_K&\equiv d\big[v^\flat\w[\i_vH-2\,c\,\i_v{*H}-c\,{*}\{\i_v{*H}\w*(\i_v{*H}\w\i_v{*H})\}]\big]=0\nn\\
%     E_{C_3}&\equiv H-2\,v^\flat\w[\i_vH-2\,c\,\i_v{*H}-c\,{*}\{\i_v{*H}\w*(\i_v{*H}\w\i_v{*H})\}]=0.
% \end{align}
\begin{align}\label{eom_Lag_np}
    &E_a \equiv  d\bigg[\frac{v^\flat \w \i_v{*H} \w \i_v{*H}+v^\flat \w \i_vH \w [\i_v{H} -4\,c\,\i_v{*H} -2\,c\,{*}\{\i_v{*H} \w *(\i_v{*H} \w \i_v{*H})\}]}{\sqrt{\i_{\pa a}(da)}}\bigg] =0\,, \nn\\
    &E_b\equiv 2\,b\,c+1=0\,, \nn\\
    &E_c\equiv\i_v{*H}\w{*}(\i_v{*H})-\tfrac{1}{4}\,\i_v{*H}\w\i_v{*H}\w{*}(\i_v{*H}\w\i_v{*H})+{*}(b^2-1)=0\,, \nn\\
    &E_A\equiv d\big[v^\flat\w[\i_vH-2\,c\,\i_v{*H}-c\,{*}\{\i_v{*H}\w*(\i_v{*H}\w\i_v{*H})\}]\big]=0\,, \nn\\
    &E_{C_3}\equiv F-v^\flat\w[\i_vH-2\,c\,\i_v{*H}-c\,{*}\{\i_v{*H}\w*(\i_v{*H}\w\i_v{*H})\}]=0\,.
\end{align}
% \begin{align}
%     E_a &\equiv  \,d\bigg[\frac{1}{\sqrt{\i_{\pa a}(da)}}(v^\flat\w\i_v{*H}\w\i_v{*H}+\i_vH\w v^\flat\w\{\i_v{H}-4\,c\,\i_v{*H}-2\,c\,{*}[\i_v{*H}\w*(\i_v{*H}\w\i_v{*H})]\})\bigg]=0 
% \end{align}
For the definition of $v^\flat$ see Appendix \ref{ext_algeb}.

${\cal L}_{np}$ is invariant under the gauge transformations $\d_d A=dY_1$ and $\d_a A=da\w Y_2\,$. The gauge symmetry associated with an infinitesimal shift of the St\"uckelberg field $a$ is
\begin{align}\label{gauge_Lag_np}
    &\d_\varphi a = \varphi\,,\quad\d_\varphi b=\frac{(2\,b\,c-1)}{2\sqrt{\i_{\pa a}(da)}}\,\i_v\i_{\pa\varphi}{*}(\i_v{*H}\w\i_v{*H})\,, \nn\\
    &\d_\varphi c=-\,\frac{2\,c^2}{\sqrt{\i_{\pa a}(da)}}\,\i_v\i_{\pa\varphi}{*}(\i_v{*H}\w\i_v{*H})\,, \quad \d_\varphi A = \frac{\varphi\,\,\i_v{\cal H}}{\sqrt{\i_{\pa a}(da)}}\,,
\end{align}
where $\,\varphi\,$ is an arbitrary scalar field and $\,{\cal H}=H-{*}\tilde{H}=H-c\,{*}[\,2\,H-\{\i_v{*H}\w*({*H}\w\i_v{*H})\}]$. By virtue of this gauge symmetry we can express the equation of motion of the St\"uckelberg field $\,a\,$ in terms of the equations of motion of the other fields in the $6D$ worldvolume of the M5-brane $-\, E_A,E_b$ and $E_c\,$:
\begin{align}\label{Ea_rel_np}
E_a=\frac{2\,E_A\w\i_v{\cal H}}{\sqrt{\i_{\pa a}(da)}}+d\bigg[
\frac{(2\,E_b-E_b^2-4\,c^2\,{*E}_c)\,v^\flat\w\i_v{*H}\w\i_v{*H}}{\sqrt{\i_{\pa a}(da)}}\bigg].
\end{align}
The above relation can be verified by observing the following:
\allowdisplaybreaks
\begin{align}\label{EAivH}
\allowdisplaybreaks
    &\quad\,\,\,\frac{2\,E_A\w\i_v{\cal H}}{\sqrt{\i_{\pa a}(da)}} \nn\\
    &=d\bigg[\frac{v^\flat}{\sqrt{\i_{\pa a}(da)}}\w\i_v{\cal H}\w\i_v{\cal H}\bigg] \nn\\
    &=d\bigg[\frac{v^\flat}{\sqrt{\i_{\pa a}(da)}}\w[\,\i_vH\w\i_vH-4\,c\,\i_vH\w\i_v{*H}+4\,c^2\,\i_v{*H}\w\i_v{*H}\nn\\
    &\qquad\quad-2\,c\,\i_vH\w{*}\{\i_v{*H}\w{*}(\i_v{*H}\w\i_v{*H})\}+4\,c^2\,\i_v{*H}\w{*}\{\i_v{*H}\w{*}(\i_v{*H}\w\i_v{*H})\}\qquad\qquad\qquad\nn\\
    &\qquad\quad+c^2\,{*}\{\i_v{*H}\w{*}(\i_v{*H}\w\i_v{*H})\}\w{*}\{\i_v{*H}\w{*}(\i_v{*H}\w\i_v{*H})\}]\bigg]
    \allowdisplaybreaks
\end{align}
\allowdisplaybreaks
and
\allowdisplaybreaks
\begin{align}
\allowdisplaybreaks
    &\qquad d\bigg[
\frac{(2\,E_b-E_b^2-4\,c^2\,{*E}_c)\,v^\flat\w\i_v{*H}\w\i_v{*H}}{\sqrt{\i_{\pa a}(da)}}\bigg] \nn\\
&=d\bigg[\frac{v^\flat}{\sqrt{\i_{\pa a}(da)}}\w\big[\,c^2\,[{*}\{\i_v{*H}\w\i_v{*H}\w{*}(\i_v{*H}\w\i_v{*H})\}]^{(0)}+(1-4\,c^2) \qquad\qquad\qquad\qquad\qquad\qquad\qquad \nn\\
&\qquad\quad-4\,c^2\,[{*}\{\i_v{*H}\w{*}(\i_v{*H})\}]^{(0)}\,\big]\,\i_v {*H}\w\i_v{*H}\bigg].\nn \allowdisplaybreaks
\end{align}
\allowdisplaybreaks
The superscript $^{(0)}$ serves as a reminder indicating that the expression inside its bracket is a scalar. Using identities \eqref{iden_Ea1} and \eqref{iden_Ea2} from Appendix \ref{app:attemptEa} the above expression becomes
\begin{align}\label{EbEc}
    &d\bigg[-\frac{c^2\,v^\flat}{\sqrt{\i_{\pa a}(da)}}\w[\,{*}\{\i_v{* H}\w{*}(\i_v{*H}\w\i_v{*H})\}+4\,\i_v {*H}\,]\w*\{\i_v {* H}\w*(\i_v{*H}\w\i_v{*H})\}\qquad\qquad\qquad\nn\\
    &\quad+\frac{(1-4\,c^2)}{\sqrt{\i_{\pa a}(da)}}\,v^\flat\w\i_v {*H}\w\i_v{*H}\bigg]\,.
\end{align}
On adding eqs. \eqref{EAivH} and \eqref{EbEc}, we get
% \begin{align}
%     d\bigg[\frac{v}{\sqrt{\i_{\pa a}(da)}}\w\big[\,\i_v{*H}\w\i_v{*H}+\i_vH\w\big\{\i_v{H}-4\,c\,\i_v{*H}-2\,c\,{*}\{\i_v{*H}\w*(\i_v{*H}\w\i_v{*H})\}\big\}\big]\bigg]\qquad
% \end{align}
\begin{align}
    d\bigg[\frac{v^\flat}{\sqrt{\i_{\pa a}(da)}} \w \big[\,\i_v{*H} \w \i_v{*H}+\i_vH \w \big\{\i_v{H}-4\,c\,\i_v{*H}-2\,c\,{*}\{\i_v{*H} \w *(\i_v{*H} \w \i_v{*H})\}\big\}\big]\bigg]\,,
\end{align}
which is the expression for $E_a$ in \eqref{eom_Lag_np}. Hence, relation \eqref{Ea_rel_np} is verified.

From relation \eqref{Ea_rel_np} we can see that whenever $\, E_A, E_b$ and $E_c\,$ are satisfied, $E_a$ is trivially satisfied. So it is evident that the St\"uckelberg field $a$ does not have independent dynamics of its own and is a gauge field.

The gauge symmetry linked with the fields $C_3$ and $C_6$ remains the same for ${\cal L}_{np}$ as for all the other Lagrangians in this paper:
\begin{align}\label{C_trans_Lag_np}
    \d_W A=-W\,,\quad \d_W C_3 =dW\,,\quad \d_W C_6 =dW\w C_3\,\, \text{where}\,\, W\,\, \text{is an arbitrary}\, 2\text{-form}.
\end{align}
This transformation commutes with the rest of the gauge transformations. All of the above gauge transformations collectively constitute a closed group.

$E_A$ in \eqref{eom_Lag_np} can be written as $d(v^\flat\w\i_v{\cal H})=0\,$ or equivalently $\,da\w d[(\i_{\pa a}{\cal H})/\i_{\pa a}(da)]=0\,$. We know that the general solution of the equation $da\wedge dU=0$, with $U$ being an arbitrary $p$-form, is $U=dX+da\wedge Y$ where $X$ and $Y$ are arbitrary $(p-1)$-forms. So the general solution to $\,da\w d[(\i_{\pa a}{\cal H})/\i_{\pa a}(da)]=0\,$ is 
\begin{align}
\frac{\i_{\pa a}{\cal H}}{\i_{\pa a}(da)}=dX+da\w Y,
\end{align}
where $X$ and $Y$ are arbitrary $1$-forms. Due to this equation we know that 
\begin{align}\label{daidacurlyH}
    da\w \i_{\pa a}{\cal H}=\i_{\pa a}(da)\,da\w dX.
\end{align}
% $da\w \i_{\pa a}{\cal H}\!=\!\i_{\pa a}(da)da\w dX \allowdisplaybreaks$.
It can be noticed that in $\i_{\pa a}{*}\tilde{H}$, $H$ always appears as $\i_{\pa a}{*H} = -\,{*}(da\w H)$ which is invariant under the gauge transformation $\d A=-\,da\w X$. So applying this gauge transformation, i.e. $\d A=-\,da\w X\,$ to eq. \eqref{daidacurlyH}, we get
\begin{align}
    da\w \i_{\pa a}{\cal H}=0\,.
\end{align}
The general solution to the above equation is $\,\i_{\pa a}{\cal H}=da\w Z\,$, with $\,Z\,$ being an arbitrary $1$-form. Taking the interior product of this equation with $\,\pa a$, we get $\,Z=({da}\w\i_{\pa a}Z)/\i_{\pa a}(da)\,$ which we plug into the relation $\,\i_{\pa a}{\cal H}=da\w Z,\,$ finally arriving at the general form of the solution for $H$:
\begin{align}
    \i_{\pa a}{\cal H}=0 \quad \Leftrightarrow \quad \i_{\pa a}H=\i_{\pa a}{*}\tilde{H} \quad \Leftrightarrow \quad da\w({*}H-\tilde H)=0.
\end{align}
It shows the nonlinear self duality of the chiral field $H$, the generalised field strength of $A$. The anti self-duality condition is obtained by redefining $\,b\rightarrow-b\,$ and $\,c\rightarrow-c\,$ in Lagrangian \eqref{Lag_np}.

% Because of this equation the following is also true:
% \begin{align}\label{da_wedge_i_daH}
%     v^\flat\w \i_v{\cal H}=v^\flat\w dX.
% \end{align}
% We apply the gauge transformation $\d A=-\,v^\flat\w X\,$ to get
% \begin{align}
%     v^\flat\w \i_v{\cal H}=0.
% \end{align}
% The general solution to the above equation is 
% \begin{align}\label{da_wedge_Z}
%     \i_v{\cal H}=v^\flat\w Z
% \end{align}
% where $Z$ is an arbitrary $1$-form. Taking the interior product of the above equation with $v$, we get, 
% \begin{align}
%     Z=v^\flat\w\i_vZ.
% \end{align}

\section{Equivalence with the PST M5-brane Lagrangian}\label{earl}

In this section we show the equivalence of the M5-brane Lagrangians given in Secs. \ref{J12} - \ref{Lag_nonpol}, with the PST M5-brane Lagrangian \cite{Pasti:1997gx,Bandos:1997ui}.
Having eliminated the auxiliary fields $B$ and $K$ in the last section, we now eliminate two more auxiliary fields $-\,b\,$ and $c\,$ such that finally we are left with only the fields $A, a, C_3$ and $C_6$ in the Lagrangian. On solving $\,E_c=0\,$ in \eqref{eom_Lag_np} we get the following two solutions for $b\,$:
\begin{align}\label{b_sol}
    b_{\pm}
    = \mp\,[1 + {*}\{\i_v{*H}\w {*(\i_v{*H})}\}
    - \tfrac{1}{4}\, 
  {*}\{\i_v{*H}\w \i_v{*H} \w {*}(\i_v{*H}\w \i_v{*H})\}]^{(1/2)}\,.
\end{align}
% \begin{align}
%     b_{\pm}
%     = \pm\,[1 + {*}\{\i_v{*H}\w {*(\i_v{*H})}\}
%     - \tfrac{1}{4}\, 
%   {*}\{\i_v{*H}\w \i_v{*H} \w {*}(\i_v{*H}\w \i_v{*H})\}]^{(\frac12)}
% \end{align}
% \begin{align}
%     b_{\pm}
%     = \pm\,\sqrt{1 + {*}\{\i_v{*H}\w {*(\i_v{*H})}\}
%     - \tfrac{1}{4}\, 
%   {*}\{\i_v{*H}\w \i_v{*H} \w {*}(\i_v{*H}\w \i_v{*H})\}}
% \end{align}
Plugging these solutions for $\,b\,$ into $E_b=0\,$ gives us
\begin{align}\label{c_sol}
    c_{\pm}=-\,\frac{1}{\,2\,b_{\pm}}
    = \pm\,\tfrac12\,[1 + {*}\{\i_v{*H}\w {*(\i_v{*H})}\}
    - \tfrac{1}{4}\, 
  {*}\{\i_v{*H}\w \i_v{*H} \w {*}(\i_v{*H}\w \i_v{*H})\}]^{(-1/2)}\,.
\end{align}
% \begin{align}
%     c_{\pm}=-\,\frac{1}{\,2\,b_{\pm}}
%     = \mp\,\frac12\,[1 + {*}\{\i_v{*H}\w {*(\i_v{*H})}\}
%     - \tfrac{1}{4}\, 
%   {*}\{\i_v{*H}\w \i_v{*H} \w {*}(\i_v{*H}\w \i_v{*H})\}]^{(-\frac12)}
% \end{align}
So we have two sets of solutions for $\,b\,$ and $\,c\,$: $\,(b_+,c_+)\,$ and $(\,b_-,c_-)\,$.
We can eliminate fields $b$ and $c\,$ from Lagrangian ${\cal L}_{np}$ by inserting either of the two solution sets into \eqref{Lag_np}. On using the solution set $\,(b_+,c_+)\,$ we get the below Lagrangian:
\begin{align}\label{Lag_e+}
  \mathcal{L}_{e+}=\,\,&\i_v{*H} \w {*}\i_vH+[4 + 4\,{*}\{\i_v{*H} \w  {*(\i_v{*H})}\}- 
  {*}\{\i_v{*H} \w  \i_v{*H}  \w  {*}(\i_v{*H} \w  \i_v{*H})\}]^{(1/2)}\,{*1}\nn\\
  & -H\w C_3-C_6\,,
\end{align}
while adopting $(\,b_-,c_-)\,$ gives
\begin{align}\label{Lag_e-}
  \mathcal{L}_{e-}=\,\,&\i_v{*H} \w {*}\i_vH-[4 + 4\,{*}\{\i_v{*H} \w  {*(\i_v{*H})}\}- 
  {*}\{\i_v{*H} \w  \i_v{*H}  \w  {*}(\i_v{*H} \w  \i_v{*H})\}]^{(1/2)}\,{*1}\nn\\
  & -H\w C_3-C_6\,.
\end{align}
The term under the square-root in the above Lagrangians is the Dirac-Born-Infeld (DBI) term which describes the dynamics of the electromagnetic fields living on the worldvolumes of D-branes and M-branes.
Lagrangian $\,{\cal L}_{e+}\,$ differs from $\,{\cal L}_{e-}\,$ by the sign of the DBI term. $\,{\cal L}_{e+}\,$ obeys the nonlinear on-shell self-duality condition $\i_vH=\i_v{*}\tilde{H}_e\,$ where
\begin{align}\label{E_K_orig}
    \tilde{H}_e=\tilde{H}|_{c=c_+}=\frac{ 2\,H-\{\i_v{*H}\w{*}({*H}\w\i_v{*H})\}}{\sqrt{4 + 4\,{*}\{\i_v{*H}\w {*(\i_v{*H})}\}
    -  
  {*}\{\i_v{*H}\w \i_v{*H} \w {*}(\i_v{*H}\w \i_v{*H})\}}}\,
\end{align}
while $\,{\cal L}_{e-}\,$ obeys the nonlinear on-shell anti self-duality condition $\i_vH=-\,\i_v{*}\tilde{H}_e$. $\,{\cal L}_{e+}\,$ is the same as the PST M5-brane Lagrangian presented earlier in \cite{Pasti:1997gx,Bandos:1997ui}. This can be seen by noting that
\begin{align}
    \det\,(g-i\,\i_v{*H})=g\,[1 + {*}\{\i_v{*H}\w {*(\i_v{*H})}\}
    - \tfrac{1}{4}\, 
  {*}\{\i_v{*H}\w \i_v{*H} \w {*}(\i_v{*H}\w \i_v{*H})\}].
\end{align}
For details see Appendix \ref{determinant}. 

% Both $\,{\cal L}_{e+}\,$ and $\,{\cal L}_{e-}\,$ are equally valid M5-brane Lagrangians, differing merely by a sign convention.
The M5-brane Lagrangians \eqref{Lag_e+}  and \eqref{Lag_e-} can be reduced to the linear-duality symmetric PST Lagrangians \cite{Pasti:1995ii,Pasti:1995tn,Pasti:1996vs}. For this we first decouple the M5-brane from the background supergravity fields $C_3$ and $C_6$, doing which eliminates the terms containing $C_3$ and $C_6$ from $\,{\cal L}_{e+}\,$ and $\,{\cal L}_{e-}\,$. Then on replacing the DBI term in $\,{\cal L}_{e+}\,$ by $\,\i_v{*F}\w{*}(\i_v{*F})\,$ we get the self-dual PST Lagrangian whereas implementing this replacement in $\,{\cal L}_{e-}\,$ gives the anti self-dual PST Lagrangian.

From here onwards we discuss the dynamics and symmetries of $\,{\cal L}_{e+}\,$. This discussion can be extended to $\,{\cal L}_{e-}\,$ by changing the signs corresponding to the sign of the DBI term. The equations of motion of $\,{\cal L}_{e+}\,$ are as follows:
\begin{align}\label{eom_orig_lag}
    &E_a\equiv d\Bigg[\frac{v^\flat\w\{\i_vH\w\i_v(H-2\,{*}\tilde{H}_e)+\i_v{*H}\w\i_v{*H}\}}{2\sqrt{\i_{\pa a}(da)}}\Bigg]=0\,,\nn\\
    &E_A\equiv d\,(v^\flat\w\i_v{\cal H}_e)=0\,, \nn\\
    &E_{C_3}\equiv F-v^\flat\w\i_v{\cal H}_e=0\,,
\end{align}
% The normalised vector field $v$ we see in $\,{\cal L}_{e+}\,$ was first introduced into a duality-symmetric Lagrangian in \cite{Khoudeir:1994zw} to make the pre-existing Lagrangian manifestly Lorentz and diffeomorphism invariant.
% \begin{align}
%     E_a&\equiv d\Bigg[\frac{v^\flat\w(\i_v{*H}\w\i_v{*H}+\i_vH\w\i_vH)}{2\sqrt{\i_{\pa a}(da)}} \nn\\
%     &\quad-\frac{v^\flat\w[2\,\i_v{*H}\w\i_vH+\i_vH\w{*}\{\i_v{*H}\w{*}(\i_v{*H}\w\i_v{*H})\}]}{2\sqrt{\i_{\pa a}(da)\,[1 + {*}\{\i_v{*H}\w {*(\i_v{*H})}\}
%     - \frac{1}{4}\, 
%   {*}\{\i_v{*H}\w \i_v{*H} \w {*}(\i_v{*H}\w \i_v{*H})\}]}}\Bigg]=0\nn\\
%     E_A&\equiv d\Bigg[v^\flat\w\i_v H-\frac{ v^\flat\w(2\,\i_v{*H}+{*}\{\i_v{*H}\w{*}(\i_v{*H}\w\i_v{*H})\})}{2\sqrt{1 + {*}\{\i_v{*H}\w {*(\i_v{*H})}\}
%     - \frac{1}{4}\, 
%   {*}\{\i_v{*H}\w \i_v{*H} \w {*}(\i_v{*H}\w \i_v{*H})\}}}\Bigg]=0\nn\\
%     E_{C_3}&\equiv F - v^\flat\w\i_v H+\frac{ v^\flat\w(2\,\i_v{*H}+{*}\{\i_v{*H}\w{*}(\i_v{*H}\w\i_v{*H})\})}{2\sqrt{1 + {*}\{\i_v{*H}\w {*(\i_v{*H})}\}
%     - \frac{1}{4}\, 
%   {*}\{\i_v{*H}\w \i_v{*H} \w {*}(\i_v{*H}\w \i_v{*H})\}}}=0
% \end{align}
where $\,{\cal H}_e=H-{*}\tilde{H}_e$. $\,{\cal L}_{e+}\,$ has the usual gauge symmetries $\,\d_d A=dY_1\,$ and $\,\d_a A=da\w Y_2\,$ with arbitrary $1$-forms $Y_1$ and $Y_2$. The infinitesimal shift symmetry of the St\"uckelberg field $a$ is now given as
\begin{align}
    \d_\varphi a = \varphi\,,\quad \d_\varphi A = \frac{\varphi\,\,\i_{\pa a}{\cal H}_e}{\i_{\pa a}(da)}\,. \label{origLgauge}
\end{align}
Here again, the above infinitesimal shift symmetry of the St\"uckelberg field $a$ enables us to see that, as expected of a gauge field, the dynamics of $a\,$ is determined by that of $A$:
\begin{align}\label{EaasEA}
E_a = \frac{E_A\w\i_{\pa a}{\cal H}_e}{\i_{\pa a}(da)}\,.
\end{align}
The fourth gauge transformation of $\,{\cal L}_{e+}\,$ involves the $11$D supergravity fields $C_3$ and $C_6$. This transformation is the same as for all the other Lagrangians in this paper; see \eqref{C_trans_Lag_np}. It can be seen from the commutators of the above transformations that they form a closed algebra.

Following a derivation similar to that shown in Sec. \ref{Lag_nonpol}, we arrive the general condition obeyed by all the solutions of $\,{\cal L}_{e+}\,$, the nonlinear self-duality condition:
\begin{align}
    \i_{\pa a}{\cal H}_e=0 \quad \Leftrightarrow \quad \i_{\pa a}H=\i_{\pa a}{*}\tilde{H}_e \quad \Leftrightarrow \quad da\w({*}H-\tilde{H}_e)=0\,.
\end{align}
Hence we have seen the duality and gauge symmetric structure of the M5-brane Lagrangian.

%Now that we have the solutions for the auxiliary fields $b$ and $c\,$ 

\section{Discussion}\label{disc}

% In the '90s Schwarz and Sen put forth a duality-symmetric action \cite{Schwarz:1993vs} which is the low-energy effective action of toroidally compactified heterotic string theory. It is not manifestly invariant under Lorentz transformations and diffeomorphism. Shortly afterwards, covariant versions of this action were presented by the PST \cite{Pasti:1995ii,Pasti:1995tn,Pasti:1996vs} via the introduction of an auxiliary vector field which was initially introduced as a constant vector field in \cite{Khoudeir:1994zw}. So the lack of manifest covariance in the earlier duality symmetric action \cite{Schwarz:1993vs} was tackled by the introduction of an auxiliary field which is also a St\"uckelberg field. However, though the PST action has most of the properties desired in a duality symmetric Lagrangian, it is not polynomial and becomes singular when the denominator containing the St\"uckelberg field vanishes. Many years after the PST action came out, it was expressed in a polynomial form in \cite{Mkrtchyan:2019opf,Bansal:2021bis}, again by introducing another auxiliary field into the Lagrangian. Similarly now, we show in this paper that the PST M5-brane Lagrangian \cite{Pasti:1997gx,Bandos:1997ui}, which is covariant by virtue of the St\"uckelberg field in it, can also be made polynomial by the introduction of a few more auxiliary fields. Hence in all of these developments, the desired characteristics of the duality symmetric action such as covariance and polynomiality are achieved by incorporating new auxiliary fields into the action.

In the '90s Schwarz and Sen put forth an electromagnetic duality symmetric action \cite{Schwarz:1993vs} which is the low-energy effective action of toroidally compactified heterotic string theory. It is not manifestly invariant under Lorentz transformations and diffeomorphism. Shortly afterwards, covariant versions of this action were presented by Pasti, Sorokin and Tonin \cite{Pasti:1995ii,Pasti:1995tn,Pasti:1996vs} via the introduction of an auxiliary vector field which was initially introduced as a constant vector field in \cite{Khoudeir:1994zw}. Subsequently the PST formalism was employed in \cite{Pasti:1997gx,Bandos:1997ui} to formulate a covariant M5-brane action. So the lack of manifest covariance in the earlier duality symmetric action \cite{Schwarz:1993vs} was tackled by the insertion of an auxiliary field which is a St\"uckelberg field. However, though the PST action has most of the properties sought in a duality symmetric Lagrangian, it is not polynomial and becomes singular when the denominator containing the St\"uckelberg field vanishes. More than two decades after the PST action came out, it was expressed in a polynomial form in \cite{Mkrtchyan:2019opf,Bansal:2021bis}, again by introducing another auxiliary field into the Lagrangian. Correspondingly now, we have shown in this paper that the non-polynomial PST M5-brane Lagrangian \cite{Pasti:1997gx,Bandos:1997ui}, which is covariant by virtue of the St\"uckelberg field $a$ in it, can also be made polynomial by the introduction of a few more auxiliary fields. Hence in all of these developments the desired characteristics of the duality symmetric action such as manifest covariance and polynomiality have been achieved by including new auxiliary fields into the action.

We have shown that the long-standing problem of the non-polynomial nature of the PST M5-brane action in \cite{Pasti:1997gx,Bandos:1997ui} can be transcended by the presence of a few additional auxiliary fields in the action. The M5-brane Lagrangian in \cite{Pasti:1997gx,Bandos:1997ui} is non-polynomial on two accounts -- the presence of the derivative of a scalar field in the denominator and the presence of the DBI term which comes with a square-root. In this paper we have presented Lagrangians in Secs. \ref{J12} and \ref{H12}, which overcome both of these non-polynomial features of the earlier Lagrangian. Owing to the structure of the Lagrangian, there are eight such polynomial Lagrangians (see Appendix \ref{alt_poly}). We have discussed the dynamics, the gauge symmetries and the (anti) self-duality conditions of our polynomial M5-brane Lagrangians.

% Focusing on the non-polynomiality due to the derivative of $a$ in the PST M5-brane Lagrangian, a crucial difference between the polynomial version of PST Lagrangian given in \cite{Bansal:2021bis} and the polynomial PST Lagrangians presented in this paper is 

% While the polynomial formulation of the PST Lagrangian \cite{Mkrtchyan:2019opf,Bansal:2021bis} has only one auxiliary field more than the original PST Lagrangian \cite{Pasti:1995ii,Pasti:1995tn,Pasti:1996vs}, the polynomial M5-brane Lagrangians given in this paper have two auxiliary fields more than in Lagrangian ${\cal L}_{np}$ \eqref{Lag_np}. 
While in the case of the PST Lagrangian \cite{Pasti:1995ii,Pasti:1995tn,Pasti:1996vs}, removing the non-polynomiality due to the derivative of the St\"uckelberg field $a$ present in the denominator requires the introduction of only one auxiliary field \cite{Mkrtchyan:2019opf,Bansal:2021bis}, removing that non-polynomiality from Lagrangian ${\cal L}_{np}$ \eqref{Lag_np}, as shown in this paper, necessitates including at least two auxiliary fields. On account of the linear self-duality condition obeyed by the PST Lagrangian ${\cal L}_{PST}=F \w {*}F+\i_v{\cal F} \w {*}\i_v({*}{\cal F})$, where ${\cal F}=(F-{*F})$, the non-polynomial term, i.e. $\i_v{\cal F} \w {*}\i_v({*}{\cal F})$, can be written as a square term (a term of the kind $T \w {*}T$) which is $-\,\i_v{\cal F} \w {*}\i_v{\cal F}$. Consequently, when making ${\cal L}_{PST}$ polynomial as in \cite{Mkrtchyan:2019opf,Bansal:2021bis}, the Lagrange multiplier is allowed to be identical with the auxiliary field whose solution we get by solving the equation of motion of the Lagrange multiplier. However, the same is not true of the non-polynomial M5-brane Lagrangian ${\cal L}_{np}$ in eq. \eqref{Lag_np}, which can be written as 
\begin{align*}
    {\cal L}_{np}&=2\,c\,H\w{*}H+\i_vH\w{*}\i_v({*}H-2\,c\,H)+\tfrac12\,c\,\i_v{*H}\w\i_v{*H}\w{*}\big(\i_v{*H}\w\i_v{*H}\big)-H\w C_3-C_6 \\
    &\quad\,+{*}(c-c\,b^2-b).
\end{align*}
Here the non-polynomial part $\,\i_vH\w{*}\i_v({*}H-2\,c\,H)+\tfrac12\,c\,\i_v{*H}\w\i_v{*H}\w{*}\big(\i_v{*H}\w\i_v{*H}\big)$ cannot be expressed as a square term due to nonlinear self-duality. So when making ${\cal L}_{np}$ polynomial, the Lagrange multiplier cannot be the same as the auxiliary field solved for by integrating out the Lagrange multiplier. Therefore, whereas the PST Lagrangian can be made polynomial by the introduction of only one auxiliary field \cite{Mkrtchyan:2019opf,Bansal:2021bis}, making ${\cal L}_{np}$ polynomial requires incorporating at least two auxiliary fields -- $B$ and $K$.

If one tries to look at the construction of the polynomial M5-brane Lagrangians presented in this paper from the perspective of starting from the PST M5-brane Lagrangian \cite{Pasti:1997gx,Bandos:1997ui}, then the following detail is worth noting. Although modifying a Lagrangian by adding a Lagrange multiplier to it and thus making it polynomial, is generally straightforward, what makes it non-trivial in the context of this work, is the specific manner of adding it which ensures that the equations of motion of the polynomial Lagrangian, when subjected to its gauge transformations, give the self-duality condition independently, i.e. without resorting to the earlier non-polynomial Lagrangian. To this end, the Lagrange multiplier along with the other auxiliary fields is required to be introduced into the non-polynomial Lagrangian in a way such that each of the following two criteria is met:
\vspace{-0.5em}
\begin{enumerate}[leftmargin=*]
    \item The equations of motion of those fields in the polynomial Lagrangian which have independent dynamics, are gauge invariant.
    \item The polynomial Lagrangian is invariant under the kind of gauge transformations which can gauge away certain parts of the general solutions of the equations of motion and their combinations (e.g. transformations \eqref{AB_trans_H12} and \eqref{B_trans_H12} in the case of ${\cal L}_{H12}$), leading to the self-duality relation.
\end{enumerate}
\vspace{-0.5em}
As the polynomial M5-brane Lagrangians presented in this paper meet both the above criteria, they form self-sufficient systems. They have all the essential symmetries and dynamics which characterise an M5-brane Lagrangian.

    % \item Substituting the solutions of the auxiliary fields into the polynomial M5-brane Lagrangian gives the non-polynomial M5-brane Lagrangian. This requirement stems from the primary characteristic of all auxiliary fields incorporated into a Lagrangian. There can be different ways of inserting auxiliary fields for meeting this requirement. It is imperative that at least one of those ways is consistent with the satisfaction of the following two criteria.

% There exist gauge transformations of the polynomial Lagrangian which can gauge away certain parts of the general solutions of the equations of motion and 
    % Presence of the particular gauge transformations which enable gauging away certain kinds of terms leading to the self-duality condition.

We have focused on the bosonic degrees of freedom coupled to bosonic background gauge fields. The local symmetry of the fermionic sector, known as the kappa-symmetry \cite{Bandos:1997ui,Simon:2011rw}, can be shown in the future. It might be worthwhile to look for the nonlinear self-interactions of M5-branes using the polynomial Lagrangians presented here. Eventually one can also venture in the direction of describing the interactions of non-abelian (2,0) superconformal field theory using this formalism.

Now that we have analytic M5-brane Lagrangians, it could be useful to construct their Hamiltonian formalism, e.g. \cite{Bergshoeff:1998vx}, which allows us to identify the first-class and second-class constraints of the theory. We can then count the number of degrees of freedom and show explicitly their correspondence with the physical fields in the theory, thus distinguishing them from the auxiliary gauge fields. The gauge system of the theory can be seen more distinctly in the Hamiltonian formalism. The polynomial form of the Lagrangians also opens up new avenues for the quantisation of the theory. Some earlier works in this direction are given in \cite{Martin:1994np,Dijkgraaf:1996hk}. A quantum treatment of Sen's duality invariant formulation is presented in \cite{Andriolo:2021gen}. We will explore the quantum formulation of the polynomial M5-brane Lagrangians in the future.

% Although the interactions of chiral fields are strongly constrained \cite{Perry:1996mk}, it has been possible to describe DBI interactions of chiral fields by a PST-like formulation in [11]. Since my M5-brane action [3] uses a new and improved version of PSTlike formulation, I intend to use our formulation to show the nonlinear self-interactions in a tractable manner.
% I would then venture in the direction of describing the interactions of non-abelian (2,0) superconformal field
% theory

% We can also think of generalising the M5-brane Lagrangian to 10 dimensions.

So far two chiral $4$-form models are known in 10 dimensions -- quadratic 4-form chiral theory along with its type IIB supergravity generalization, and a generalisation of  Bialynicki-Birula electrodynamics \cite{Buratti:2019guq,Bandos:2020hgy}. However, an M5-brane-like action has not been generalised to 10 dimensions as it requires solving the condition imposed by the PST gauge invariance. It has proven difficult to solve that condition in 10 and higher dimensions, even perturbatively. Our new polynomial formalism seems favourable for addressing this problem.

As the origin of many D-branes in string theory is the M5-brane in M-theory, dimensionally reducing an M5-brane action gives us D-brane actions. For e.g. performing double dimensional reduction of an M5-brane on a circle gives the world volume dual of the D4-brane of type IIA string theory \cite{Aganagic:1997zk} while doing that on a torus gives the D3-brane of type IIB string theory directly reduced on a circle \cite{Berman:1998va}. A D3-brane action was also obtained in \cite{Nurmagambetov:1998gp} by dimensionally reducing the PST M5-brane action. Now the polynomial Lagrangians given in this paper can also be used to implement dimensional reductions giving D-brane actions. Thus, in a variety of ways the polynomial structure of these M5-brane Lagrangians may prove to be relatively more tractable for carrying out further developments with M5-brane actions.

\vspace{2em}

% \section*{Acknowledgements}

\textbf{\large Acknowledgements}

\vspace{0.25em}

The author is indebted to Pichet Vanichchapongjaroen for his useful input during the early stages of this project. In addition, the author is grateful to Karapet Mkrtchyan for suggesting this research topic. A special thanks goes to Oleg Evnin, Carlo Iazeolla, Akash Jain, Silvia Nagy, Valdo Tatitscheff and Ivonne Zavala for their useful discussions and feedback.

\vspace{2em}

\appendix

\addtocontents{toc}{\protect\setcounter{tocdepth}{1}}

\section{Alternate Polynomial M5-brane Lagrangians}\label{alt_poly}

The polynomial Lagrangians presented in sections \ref{J12} and \ref{H12} have two terms containing both $c$ and $K$. By changing the number of $J\,$ (or $H$)$\,$ fields in those terms, we can get $8$ different Lagrangians: ${\cal L}_{00},{\cal L}_{01},{\cal L}_{02},{\cal L}_{03},{\cal L}_{10},{\cal L}_{11},{\cal L}_{12},{\cal L}_{13}$. Here we show four of them: ${\cal L}_{00},{\cal L}_{02},{\cal L}_{11},{\cal L}_{13}$. The remaining Lagrangians can be similarly constructed and analysed.

%particular terms with the coefficient $c$, one of which is quadratic in $2/3$-form fields and another is quartic.
% Here we show a couple of different polynomial M5-brane Lagrangians all of which are manifestly covariant. The distinction between them is mainly with respect to the second quadratic terms and the quartic terms, which carry a different number of $J\,$ (or $H$)$\,$ fields.

\subsection{Lagrangian ${\cal L}_{00}$}

% hide subsections in the toc:

\subsubsection{Lagrangian ${\cal L}_{J00}$}

\begin{align}
    \mathcal{L}_{J00}=\,\,& da\w K\w[{*}J-2\,c\,\i_{\pa a}{*K}+\tfrac12\,c\,\i_{\pa a}(da)\,K\w\i_{\pa a}{*}(K\w K)]+ H\w(J-C_3) \nn\\
    &+2\,{*}(c-c\,b^2-b)\,.
\end{align}
Equations of motion:
\begin{align}
    &E_a\equiv d\big[K\w[ {*J}-4\,c\,\i_{\pa a}{*K}+\{da+\i_{\pa a}(da)\}\,c\,K\w\i_{\pa a}{*}(K\w K)]\big]+G\w J=0\,,\nn\\
    &E_b\equiv 2\,b\,c+1=0\,, \nn\\
    &E_c\equiv da\w K\w[\,2\,\i_{\pa a}{*K}-\tfrac12\,\i_{\pa a}(da)\,K\w\i_{\pa a}{*}(K\w K)]+2\,{*}(b^2-1)=0\,, \nn\\
    &E_A\equiv d(\i_{\pa a}{*K}-H+a\,G)=0 \,,\nn\\
    &E_B\equiv d[ a\,(\i_{\pa a}{*K}-H)]=0 \,, \nn\\
    &E_K\equiv da\w[{*}J-4\,c\,\i_{\pa a}{*K}+2\,c\,\i_{\pa a}(da)\,K\w\i_{\pa a}{*}(K\w K)]=0 \,,\nn\\
    &E_{C_3}\equiv \i_{\pa a}{*K}+F+a\,G=0 \,.
\end{align}
On subjecting the general solution to the equation $E_A-d[\i_{\pa a}(E_B-a\,E_A)/\i_{\pa a}(da)]=0,$ to the gauge transformations $\d_d B= dY_1$ and $\d A=-\,a\,da\w Y_2\,,\d B=da\w Y_2$ with appropriate choices for $Y_1$ and $Y_2$, we get the nonlinear self-duality condition as follows:
\begin{align}\label{sdual_J00}
    \i_{\pa a}J=-\,2\,\{\i_{\pa a}(da)\}B.
\end{align}
% ${*}da\w\i_{\pa a}(E_B-a\,E_A)=0\,$ gives
% \begin{align}\label{Ksol_J00}
%     K=\frac{\i_{\pa a}{*}J}{\i_{\pa a}(da)}\,,
% \end{align}
% and it can be checked that
% \begin{align}\label{Bsol_J00}
%     B&=\frac{\i_{\pa a}J}{2\,\i_{\pa a}(da)}-2\,c\,K+c\,{*}[da\w K\w{*}(da\w K\w K)]\,.
% \end{align}
% Substituting the expressions for $K$ and $B$ from the above two equations into eq. \eqref{sdual_J00}, we obtain
% \begin{align}\label{selfdual_J00}
%     \i_{\pa a}J=\i_{\pa a}{*}\tilde{J},\quad \text{where}\,\,\tilde{J}=2\,c\,J-\frac{\,c\,\i_{\pa a}{*J}\w*({*J}\w\i_{\pa a}{*J})\,}{\i_{\pa a}(da)}\,.
% \end{align}
% This implies that the solution for $B$ in \eqref{Bsol_J00} vanishes on-shell. Hence we get the nonlinear self-duality condition $\i_{\pa a}H=\i_{\pa a}{*}{\tilde H}$ where $\tilde{H}=2\,c\,H-[\,c\,\i_{\pa a}{*H}\w*({*H}\w\i_{\pa a}{*H})\,]/\i_{\pa a}(da)$.

\subsubsection{Lagrangian ${\cal L}_{H00}$}

Redefining $A\rightarrow (A-a\,B)\,$ in ${\cal L}_{J00}$ we arrive at ${\cal L}_{H00}$:
\begin{align}
    \mathcal{L}_{H00}=\,\,&da\w K\w[{*}H-\i_{\pa a}{*}(2\,c\, K+B)+\tfrac12\,c\,\i_{\pa a}(da)\, K\w\i_{\pa a}{*}( K\w K)] -H\w(da\w B+C_3)\nn\\
    &-C_6+2\,{*}(c-c\,b^2-b)\,. 
\end{align}
Equations of motion:
\begin{align}
    &E_a\equiv d[K\w({*}H-\i_{\pa a}{*}(4\,c\,da\w K+B) +2\,c\,\i_{\pa a}(da)\, K\w K\w\i_{\pa a}{*}(K\w K)\nn\\
    &\qquad\quad -c\,da\w\i_{\pa a}(K\w K)\w\i_{\pa a}{*}(K\w K)+B\w(H-\i_{\pa a}{*}K)]=0\,, \nn\\
    &E_b\equiv 2\,b\,c+1=0\,, \nn\\
    &E_c\equiv da\w K\w[\,2\,\i_{\pa a}{*}K-\tfrac12\,\i_{\pa a}(da)\, K\w\i_{\pa a}{*}( K\w K)]+2\,{*}(b^2-1)=0\,, \nn\\
    &E_A\equiv d(\i_{\pa a}{*K}-H-da\w B)=0\,,\nn\\
    &E_B\equiv da\w(\i_{\pa a}{*K}-H)=0\,, \nn\\
    &E_K\equiv da\w[{*}H-\i_{\pa a}{*}(4\,c\, K+B)+2\,c\,\i_{\pa a}(da)\, K\w\i_{\pa a}{*}( K\w K)]=0\,, \nn\\
    &E_{C_3}\equiv \i_{\pa a}{*K}+F-da\w B=0\,.
\end{align}
Applying the gauge transformations $\d A=da\w Z_1\,, \d B=-\,dZ_1$ and $\d_a B=da\w Z_2$ with suitable choices for $Z_1$ and $Z_2$, on the general solution to the equation $E_A-d[\i_{\pa a}E_B/\i_{\pa a}(da)]=0$, gives
\begin{align}\label{sdual_H02}
    \i_{\pa a}H=-\,\{\i_{\pa a}(da)\} B\,.
\end{align}
$E_B=0$ is satisfied by the following solution for $K$:
\begin{align}\label{Ksol_H00}
    K=\frac{\i_{\pa a}{*}H}{\i_{\pa a}(da)}\,,
\end{align}
and $E_K=0$ admits the solution:
\begin{align}\label{Bsol_H00}
    B&=\frac{\i_{\pa a}H}{\i_{\pa a}(da)}-4\,c\,K+2\,c\,\i_{\pa a}{*}[K\w\i_{\pa a}{*}(K\w K)]\,.
\end{align}
Drawing on solutions \eqref{Ksol_H00} and \eqref{Bsol_H00}, we can see that eq. \eqref{sdual_H02} gives the same nonlinear self-duality condition as in eq. \eqref{selfdual-cond}, i.e.
\begin{align}
    \i_{\pa a}H=\i_{\pa a}{*}\tilde{H},\quad \text{where}\,\,\tilde{H}=2\,c\,H-\frac{\,c\,\i_{\pa a}{*H}\w{*}({*H}\w\i_{\pa a}{*H})\,}{\i_{\pa a}(da)}.
\end{align}

% \addtocontents{toc}{\protect\setcounter{tocdepth}{2}}

\subsection{Lagrangian ${\cal L}_{02}$}

% hide subsections in the toc:
\subsubsection{Lagrangian ${\cal L}_{J02}$}

\begin{align}
    \mathcal{L}_{J02}
    = \,\,& da\w K\w [ {*}J-2\,c\,\i_{\pa a}{*K}+\tfrac12\,c\,K\w{*}({*}J\w\i_{\pa a}{*J})]+H\w(J-C_3)-C_6\nn\\
    &+2\,{*}(c-c\,b^2-b)\,.
\end{align}

Equations of motion:
\begin{align}
    &E_a\equiv d[K\w\{{*}J-2\,c\,J+c\,K\w{*}({*}J\w\i_{\pa a}{*J})\}]+G\w(J-\i_{\pa a}{*K}-2\,c\,da\w K) \nn\\
    &\qquad\quad-c\,da\w K\w K\w{*}(\i_{\pa a}{*G}\w\i_{\pa a}{*J})=0\,,\nn\\
    &E_b\equiv 2\,b\,c+1=0 \nn\\
    &E_c\equiv da\w K\w [2\,\i_{\pa a}{*K}-\tfrac12\,K\w{*}({*}J\w\i_{\pa a}{*J})]+2\,{*}(b^2-1)=0\,, \nn\\
    &E_A\equiv d[\,\i_{\pa a}{*K}-H-c\,{*}\{\i_{\pa a}{*}(K\w K)\w\i_{\pa a}{*J}\}+a\,G]=0\,,\nn\\
    &E_B\equiv d\big[\,a\,[\i_{\pa a}{*K}-H-c\,{*}\{\i_{\pa a}{*}(K\w K)\w\i_{\pa a}{*J}\}]\big]=0\,, \nn\\
    &E_K\equiv da\w [ {*}J-4\,c\,\i_{\pa a}{*K}+c\,K\w{*}({*}J\w\i_{\pa a}{*J})]=0\,, \nn\\
    &E_{C_3} \equiv \i_{\pa a}{*K}+F-c\,{*}\{\i_{\pa a}{*}(K\w K)\w\i_{\pa a}{*J}\}+a\,G=0\,.
\end{align}
In view of two gauge symmetries of ${\cal L}_{J02}$ viz. $\d_d B= dY_1$ and $\d A=-\,a\,da\w Y_2\,$ alongside $\d B=da\w Y_2,$ the equation $E_A-d[\i_{\pa a}(E_B-a\,E_A)/\i_{\pa a}(da)]=0$ leads to the following:
\begin{align}\label{selfdual_J02}
    \i_{\pa a}J=\{\i_{\pa a}(da)\}[\,-\,2\,B+c\,{*}\{{*}(K\w K)\w\i_{\pa a}{*J}\}]\,.
\end{align}
${*}[da\w\i_{\pa a}(E_B-a\,E_A)]=0\,$ effectively gives
\begin{align}\label{Ksol_J02}
    K=\frac{\i_{\pa a}{*}J}{\i_{\pa a}(da)}\,,
\end{align}
Plugging this expression for $K$ into eq. \eqref{selfdual_J02} gives us the nonlinear self-duality condition as below:
\begin{align}
    \i_{\pa a}J=\frac{c\,{*}[{*}(\i_{\pa a}{*J}\w \i_{\pa a}{*J})\w\i_{\pa a}{*J}]}{\i_{\pa a}(da)}-2\,\{\i_{\pa a}(da)\} B.
\end{align}

\subsubsection{Lagrangian ${\cal L}_{H02}$}

\begin{align}
    \mathcal{L}_{H02}=\,\,& da\w K\w[{*H}-\i_{\pa a}{*}(2\,c\, K+ B)+\tfrac12\,c\,K\w{*}({*H}\w\i_{\pa a}{*H})]-H\w(da\w B+C_3) \nn\\
    &-C_6+2\,{*}(c-c\,b^2-b)\,.
\end{align}

Equations of motion:
\begin{align}
    &E_a\equiv d[K\w\{{*H}-\i_{\pa a}{*}(4\,c\,K+B)\} -c\,H\w{*}\{{*}(K\w K)\w\i_{\pa a}{*H}\}+B\w(H-\i_{\pa a}{*K})]=0\,, \nn\\
    &E_b\equiv 2\,b\,c+1=0\,, \nn\\
    &E_c\equiv da\w K\w[\,2\,\i_{\pa a}{*K}- \tfrac{1}{2}\, K\w{*}({*}H\w\i_{\pa a}{*H})]+2\,{*}(b^2-1)=0\,, \nn\\
    &E_A\equiv d\big[\,\i_{\pa a}{*K}-H+da\w[ c\,{*}\{{*}(K\w K)\w\i_{\pa a}{*H}\}-B\,]\big]=0\,, \nn\\
    &E_B\equiv da\w(\i_{\pa a}{*K}-H)=0\,, \nn\\
    &E_K\equiv da\w[{*H}-\i_{\pa a}{*}(4\,c\, K+ B)+c\,K\w{*}({*H}\w\i_{\pa a}{*H})]=0\,, \nn\\
    &E_{C_3} \equiv \i_{\pa a}{*K}+F+da\w[\,c\,{*}\{{*}(K\w K)\w\i_{\pa a}{*H}\}-B\,]=0\,.
\end{align}
The equation $E_A-d[\i_{\pa a}E_B/\i_{\pa a}(da)]=0$ modulo the terms that can be gauged away via the gauge transformation $\d A=da\w Z_1\,$ with $\d B=-\,dZ_1,$ and $\d_a B=da\w Z_2$, gives
\begin{align}\label{selfdual_H02}
    \i_{\pa a}H=\{\i_{\pa a}(da)\}[-B+c\,{*}\{{*}(K\w K)\w\i_{\pa a}{*H}\}]\,.
\end{align}
Neglecting the terms that can be gauged away, the solutions for $K$ and $B$ are
\begin{align}\label{Bsol_H02}
    K&=\frac{\i_{\pa a}{*}H}{\i_{\pa a}(da)}\,,\nn\\
    B &=\frac{\i_{\pa a}H}{\i_{\pa a}(da)}-4\,c\,K-\frac{c}{\i_{\pa a}(da)}\,{*}[K\w{*}(\i_{\pa a}{*}H\w\i_{\pa a}{*}H)]\,.
\end{align}
On inserting these solutions for $B$ and $K$ into relation \eqref{selfdual_H02}, we get the usual nonlinear self-duality condition \eqref{selfdual-cond}, i.e. $\i_{\pa a}H=\i_{\pa a}{*}\tilde{H}$ where $\tilde{H}=2\,c\,H-[\,c\,\i_{\pa a}{*H}\w*({*H}\w\i_{\pa a}{*H})/\i_{\pa a}(da)]$.

% {\protect\setcounter{tocdepth}{3}}

\subsection{Lagrangian ${\cal L}_{11}$}

% hide subsections in the toc:
\subsubsection{Lagrangian ${\cal L}_{J11}$}

\begin{align}\label{Lag_J11}
    \mathcal{L}_{J11} = da\w K\w[{*}J-2\,c\, J+\tfrac12\,c\,K\w\i_{\pa a}{*}(K\w\i_{\pa a}{*}J)] +H \w (J-C_3)-C_6 +2\,{*}(c-c\,b^2-b).
\end{align}

Equations of motion:
\begin{align}
    &E_a\equiv d[K\w({*}J-2\,c\,J)+\tfrac12\,c\,K\w \{K\w\i_{\pa a}{*}(K\w\i_{\pa a}{*}H)+\i_{\pa a}{*H}\w\i_{\pa a}{*}(K\w K)\}\nn\\
    &\qquad\quad+\tfrac12\,c\,H\w\i_{\pa a}{*}\{K\w\i_{\pa a}{*}(K\w K)\}]+G\w(J-\i_{\pa a}{*K}-2\,c\,da\w K) \nn\\
    &\qquad\quad-\tfrac12\,c\,da\w K\w K\w\i_{\pa a}{*}(K\w\i_{\pa a}{*G})=0\,, \nn\\
    &E_b\equiv 2\,b\,c+1=0 \,,\nn\\
    &E_c\equiv K\w[\,2\,da\w J- \tfrac12\,da\w K\w\i_{\pa a}{*}(K\w\i_{\pa a}{*}J)]+2\,{*}(b^2-1)=0\,, \nn\\
    &E_A\equiv d\big[\i_{\pa a}{*}K-H+da\w[\,2\,c\, K-\tfrac12\,c\,\i_{\pa a}{*}\{K\w\i_{\pa a}{*}(K\w K)\}]+a\,G\big]=0\,, \nn\\
    &E_B\equiv d\big[a\,\{\i_{\pa a}{*}K-H+da\w[\,2\,c\, K-\tfrac12\,c\,\i_{\pa a}{*}\{K\w\i_{\pa a}{*}(K\w K)\}\big]=0\,, \nn\\
    &E_K\equiv da\w[\,{*J}-2\,c\,J+c\,da\w\{ K\w\i_{\pa a}{*}(K\w\i_{\pa a}{*}J)+\tfrac12\,\i_{\pa a}{*J}\w\i_{\pa a}{*}(K\w K)\}]=0\,, \nn\\
    &E_{C_3}\equiv \i_{\pa a}{*} K+F+da\w[\,2\,c\, K-\tfrac12\,c\,\i_{\pa a}{*}\{K\w\i_{\pa a}{*}(K\w K)\}]+a\,G=0\,.
\end{align}
By removing the terms allowable to be gauged away, the equation $E_A-d[\i_{\pa a}(E_B-a\,E_A)/\i_{\pa a}(da)]=0$ gives us the following relation:
\begin{align}\label{selfdual_breveJ12}
    \i_{\pa a}J=\{\i_{\pa a}(da)\}[\,2\,(c\,K-B)-\tfrac12\,c\,\i_{\pa a}{*}\{K\w\i_{\pa a}{*}(K\w K)\}]\,.
\end{align}
Also, from ${*}[da\w\i_{\pa a}(E_B-a\,E_A)]=0\,$ we effectively get
\begin{align}\label{Ksol_breveJ12}
    K=\frac{\i_{\pa a}{*}J}{\i_{\pa a}(da)}.
\end{align}
So eq. \eqref{selfdual_breveJ12} becomes
\begin{align}
    \i_{\pa a}J=2\,\i_{\pa a}{*}J+\frac{c\,{*}[\i_{\pa a}{*J}\w{*}(\i_{\pa a}{*J})\w\i_{\pa a}{*J}]}{\i_{\pa a}(da)}-2\,\{\i_{\pa a}(da)\}B,
\end{align}
the nonlinear self-duality condition of ${\cal L}_{J11}\,$.

\subsubsection{Lagrangian ${\cal L}_{H11}$}

\begin{align}\label{Lag_H11}
    \mathcal{L}_{H11}
    =\,\,& da\w K\w[{*H}-2\,c\,H-\i_{\pa a}{*}B+\tfrac12\,c\,K\w\i_{\pa a}{*}(K\w\i_{\pa a}{*}H)]-H\w(da\w B+C_3)-C_6\nn\\
    &+2\,{*}(c-c\,b^2-b)\,.
\end{align}
Equations of motion:
\begin{align}
    &E_a\equiv d[K\w({*H}-2\,c\,H-\i_{\pa a}{*}B) +\tfrac12\,c\,K\w \{K\w\i_{\pa a}{*}(K\w\i_{\pa a}{*}H)+\i_{\pa a}{*H}\w\i_{\pa a}{*}(K\w K)\}\nn\\
    &\qquad\quad+\tfrac12\,c\,H\w\i_{\pa a}{*}\{K\w\i_{\pa a}{*}(K\w K)\}] +B\w(H-\i_{\pa a}{*} K)]=0\,, \nn\\
    &E_b\equiv 2\,b\,c+1=0\,, \nn\\
    &E_c\equiv da\w K\w[\,2\, H- \tfrac12\,K\w\i_{\pa a}{*}(K\w\i_{\pa a}{*}H)]+2\,{*}(b^2-1)=0\,, \nn\\
    &E_A\equiv d\big[\i_{\pa a}{*}K-H+da\w[\,2\,c\, K-\tfrac12\,c\,\i_{\pa a}{*}\{K\w\i_{\pa a}{*}(K\w K)\}-B\,]\big]=0\,, \nn\\
    &E_B\equiv da\w(\i_{\pa a}{*K}-H)=0\,, \nn\\
    &E_K\equiv da\w[{*H}-2\,c\,H-\i_{\pa a}{*}B+c\,da\w\{ K\w\i_{\pa a}{*}(K\w\i_{\pa a}{*}H)+\tfrac12\,\i_{\pa a}{*H}\w\i_{\pa a}{*}(K\w K)\}]=0\,, \nn\\
    &E_{C_3}\equiv \i_{\pa a}{*} K+F+da\w[\,2\,c\, K-\tfrac12\,c\,\i_{\pa a}{*}\{K\w\i_{\pa a}{*}(K\w K)\}-B\,]=0\,.
\end{align}
The particular combination of $E_A$ and $E_B,$ $E_A-d[\i_{\pa a}E_B/\i_{\pa a}(da)]=0,$ when subjected to two of the gauge transformations of ${\cal{L}}_{H11}$, enables us to see that
\begin{align}\label{selfdual_breveH12}
    \i_{\pa a}H=\{\i_{\pa a}(da)\}[\,2\,c\,K-B-\tfrac12\,c\,\i_{\pa a}{*}\{K\w\i_{\pa a}{*}(K\w K)\}]\,.
\end{align}
The fields $K$ and $B$ have the following solutions (shown without the terms which are redundant on account of gauge symmetries) :
\begin{align}\label{Bsol_Htilde12}
    K&=\frac{\i_{\pa a}{*}H}{\i_{\pa a}(da)},\nn\\
    B&=\frac{\i_{\pa a}H-2\,c\,\i_{\pa a}{*H}+c\,\i_{\pa a}{*}\{K\w\i_{\pa a}{*}(K\w\i_{\pa a}{*}H)+\tfrac12\,\i_{\pa a}{*H}\w\i_{\pa a}{*}(K\w K)\}}{\i_{\pa a}(da)}\,.
\end{align}
Again, it can be seen that substituting solutions \eqref{Bsol_Htilde12} into eq. \eqref{selfdual_breveH12} gives us the usual nonlinear self-duality condition \eqref{selfdual-cond} in terms of the fields $a$ and $H$: $\i_{\pa a}H=\i_{\pa a}{*}\tilde{H}$.

\subsection{Lagrangian ${\cal L}_{13}$}

% hide subsections in the toc:
% \addtocontents{toc}{\protect\setcounter{tocdepth}{2}}

\subsubsection{Lagrangian ${\cal L}_{J13}$}

\begin{align}
    \mathcal{L}_{J13}=\,\,&da\w K\w[\{\i_{\pa a}(da)\}(*J-2\,c\,J)+\tfrac12\,c\,\i_{\pa a}{*J}\w{*}({*}J\w\i_{\pa a}{*J})]+H\w(J-C_3)-C_6 \nn\\
    &+2\,{*}(c-c\,b^2-b)\,.
\end{align}
Equations of motion:
\begin{align}\label{eom_J13}
    &E_a\equiv d[\{\i_{\pa a}(da)\}K\w\{{*}J-2\,c\,J+\tfrac12\,c\,\i_{\pa a}{*J}\w{*}({*}J\w\i_{\pa a}{*J})\}+\i_{\pa a}\{ K\w{*}\i_{\pa a}(J-2\,c\,{*}J)\}\nn\\
    &\qquad\quad-\tfrac12\,c\,J\w{*}\{K\w{*}(\i_{\pa a}{*J}\w\i_{\pa a}{*J})-\tfrac12\,c\,J\w{*}\{{*J}\w\i_{\pa a}{*}(K\w\i_{\pa a}{*J})]\nn\\
    &\qquad\quad+G\w\{J-\{\i_{\pa a}(da)\}(\i_{\pa a}{*K}-2\,c\,da\w K)\}-c\,K\w[\,\tfrac12\,\{\i_{\pa a}{*G}\w{*}(\i_{\pa a}{*J}\w\i_{\pa a}{*J})\nn\\
    &\qquad\quad+\i_{\pa a}{*J}\w{*}(\i_{\pa a}{*J}\w\i_{\pa a}{*G})]=0\,, \nn\\
    &E_b\equiv 2\,b\,c+1=0\,, \nn\\
    &E_c\equiv da\w K\w[\,2\,\{\i_{\pa a}(da)\}\,J-\tfrac12\,\i_{\pa a}{*J}\w{*}({*}J\w\i_{\pa a}{*J})]+2\,{*}(b^2-1)=0\,, \nn\\
    &E_A\equiv d[\,\{\i_{\pa a}(da)\}(\i_{\pa a}{*}K+2\,c\,da\w K)-H-\tfrac12\,c\,{*}\i_{\pa a}\{K\w{*}(\i_{\pa a}{*J}\w\i_{\pa a}{*J})\}\nn\\
    &\qquad\quad-c\,{*}\{\i_{\pa a}{*J}\w\i_{\pa a}{*}(K\w\i_{\pa a}{*J})\}+a\,G\,]=0\,, \nn\\
    &E_B\equiv d\big[\,a\,[\{\i_{\pa a}(da)\}(\i_{\pa a}{*}K+2\,c\,da\w K)-H-\tfrac12\,c\,{*}\i_{\pa a}\{K\w{*}(\i_{\pa a}{*J}\w\i_{\pa a}{*J})\}\nn\\
    &\qquad\quad-c\,{*}\{\i_{\pa a}{*J}\w\i_{\pa a}{*}(K\w\i_{\pa a}{*J})\}]\,\big]=0\,, \nn\\
    &E_K\equiv da\w[\{\i_{\pa a}(da)\}(*J-2\,c\,J)+\tfrac12\,c\,\i_{\pa a}{*J}\w{*}({*}J\w\i_{\pa a}{*J})]=0\,, \nn\\
    &E_{C_3}\equiv \{\i_{\pa a}(da)\}(\i_{\pa a}{*}K+2\,c\,da\w K)+F-\tfrac12\,c\,{*}\i_{\pa a}\{K\w{*}(\i_{\pa a}{*J}\w\i_{\pa a}{*J})\}\nn\\
    &\qquad\quad-c\,{*}\{\i_{\pa a}{*J}\w\i_{\pa a}{*}(K\w\i_{\pa a}{*J})\}+a\,G=0\,.
\end{align}
We consider the equation $E_A-d[\i_{\pa a}(E_B-a\,E_A)/\i_{\pa a}(da)]=0,$ which when subjected to the gauge transformations $\d_d B= dY_1$ and $\d A=-\,a\,da\w Y_2\,$ alongside $\d B=da\w Y_2,$ provides us with the following relation:
\begin{align}\label{selfdual_J13}
    \i_{\pa a}J=\{\i_{\pa a}(da)\}\big[2\,[\{\i_{\pa a}(da)\}\,c\, K\!-\!B]\!+\!\tfrac12\,c\,{*}\{K\w{*}({\i_{\pa a}{*J}}\w\i_{\pa a}{*J})\}\!+\!c\,{*}\{\i_{\pa a}{*J}\w{*}(K\w\i_{\pa a}{*J})\}\big].
\end{align}
${*}[da\w\i_{\pa a}(E_B-a\,E_A)]=0\,$ effectively gives
\begin{align}\label{Ksol_J00}
    K=\frac{\i_{\pa a}{*}J}{\{\i_{\pa a}(da)\}^2}.
\end{align}
% and it can be checked that
% \begin{align}
%     B=\frac{\,c\,\i_{\pa a}{*J}\w*(\i_{\pa a}{*J}\w\i_{\pa a}{*J})\,}{2\,\i_{\pa a}(da)}.
% \end{align}
Inserting the above expression for $K$ into eq. \eqref{selfdual_J13}, the nonlinear self-duality relation of ${\cal L}_{J13}$ gets expressed as
\begin{align}
    \i_{\pa a}J=2\,c\,\i_{\pa a}{*}J+\frac{3\,c\,\i_{\pa a}{*J}\w{*}(\i_{\pa a}{*J}\w\i_{\pa a}{*J})\,}{2\,\i_{\pa a}(da)}-2\,\{\i_{\pa a}(da)\}B.
\end{align}

\subsubsection{Lagrangian ${\cal L}_{H13}$}

\begin{align}
    \mathcal{L}_{H13}=\,\,&da\w K\w[\{\i_{\pa a}(da)\}({*}H-2\,c\,H-\i_{\pa a}{*B})+\tfrac12\,c\,\i_{\pa a}{*H}\w{*}\{{*H}\w\i_{\pa a}{*H}\}]\nn\\
    &-H\w(da\w B+C_3)-C_6+2\,{*}(c-c\,b^2-b).
\end{align}
Equations of motion:
\begin{align}
    &E_a\equiv d[\,3\,\{\i_{\pa a}(da)\}K\w({*}H-2\,c\,H-\i_{\pa a}{*B})-2\,da\w\i_{\pa a}\{K\w({*}H-2\,c\,H-\i_{\pa a}{*B})\}\nn\\
    &\qquad\quad-\tfrac12\,c\,H\w{*}\{K\w{*}(\i_{\pa a}{*H}\w\i_{\pa a}{*H})\}-c\,H\w{*}\{\i_{\pa a}{*H}\w{*}(K\w\i_{\pa a}{*H})\}\nn\\
    &\qquad\quad+B\w\{H-\{\i_{\pa a}(da)\}\i_{\pa a}{*K}\}]=0\,,\nn\\
    &E_b\equiv 2\,b\,c+1=0\,, \nn\\
    &E_c\equiv da\w K\w[\,2\,\{\i_{\pa a}(da)\} H-\tfrac12\,\i_{\pa a}{*H}\w{*}({*H}\w\i_{\pa a}{*H})]+2\,{*}(b^2-1)=0\,, \nn\\
    &E_A\equiv d\big[\{\i_{\pa a}(da)\} \i_{\pa a}{*K}-H+da\w[\,2\,\{\i_{\pa a}(da)\}\,c\, K+\tfrac12\,c\,{*}\{K\w{*}({\i_{\pa a}{*H}}\w\i_{\pa a}{*H})\}\nn\\
    &\qquad\quad+c\,{*}\{\i_{\pa a}{*H}\w{*}(K\w\i_{\pa a}{*H})\}-B\,]\big]=0\,,\nn\\
    &E_B\equiv da\w[\{\i_{\pa a}(da)\}\i_{\pa a}{*K}-H]=0\,, \nn\\
    &E_K\equiv da\w[\{\i_{\pa a}(da)\}({*}H-2\,c\,H-\i_{\pa a}{*B})+\tfrac12\,c\,\i_{\pa a}{*H}\w{*}({*H}\w\i_{\pa a}{*H})]=0\,,\nn\\
    &E_{C_3}\equiv \{\i_{\pa a}(da)\} \i_{\pa a}{*K}+F+da\w[\,2\,c\, K+\tfrac12\,c\,{*}\{K\w{*}({\i_{\pa a}{*H}}\w\i_{\pa a}{*H})\}\nn\\
    &\qquad\quad+c\,{*}\{\i_{\pa a}{*H}\w{*}(K\w\i_{\pa a}{*H})\}-B\,]=0\,.
\end{align}
We apply the gauge transformations $\d A=da\w Z_1\,, \d B=-\,dZ_1$ and $\d_a B=da\w Z_2$ to the general solution of the equation $E_A-d[\i_{\pa a}E_B/\i_{\pa a}(da)]=0$, and get
\begin{align}\label{sdualH13}
    \i_{\pa a}H\!=\!\{\i_{\pa a}(da)\}\big[[2\,c\,\{\i_{\pa a}(da)\}K\!-\!B]\!+\!\tfrac{c}{2}{*}\{K\w{*}({\i_{\pa a}{*H}}\w\i_{\pa a}{*H})\}\!+\!c\,{*}\{\i_{\pa a}{*H}\w{*}(K\w\i_{\pa a}{*H})\}\big].
\end{align}
On solving $E_B=0$ and $E_K=0$ we find,
\begin{align}\label{Bsol_H13}
    K&=\frac{\i_{\pa a}{*}H}{\{\i_{\pa a}(da)\}^2}\,,\nn\\	B&=\frac{\i_{\pa a}(H-2\,c\,{*H})}{\i_{\pa a}(da)}-\frac{c\,{*}\{\i_{\pa a}{*H}\w{*}(\i_{\pa a}{*H}\w\i_{\pa a}{*H})\}}{2\,\{\i_{\pa a}(da)\}^2}\,.
\end{align}
Here the solutions for $K$ and $B$ are independent of each other. $B$ and $K$ can be eliminated from eq. \eqref{sdualH13} using the above solutions, arriving at the nonlinear self-duality condition:
$$\i_{\pa a}H=\i_{\pa a}{*}\tilde{H},\quad \text{where}\,\,\tilde{H}=2\,c\,H-\frac{\,c\,\i_{\pa a}{*H}\w{*}({*H}\w\i_{\pa a}{*H})\,}{\i_{\pa a}(da)}.$$

% We could also have ${\cal L}_{03}$.

\section{Notation and Conventions}\label{convDF}

\subsection{Levi-Civita Tensor and Generalised Kronecker Delta Function}\label{Levciv}

The Levi-Civita symbol, denoted by $\hat\varepsilon_{\m_1\m_2...\m_d}$, is defined as follows:
\begin{align}\label{LeviCivitensdens}
    \hat\varepsilon^{\m_1\m_2...\m_d}= \begin{cases} +1\,\, \text{if}\,\, (\m_1, \m_2, ..., \m_d)\,\, \text{is an even permutation of}\,\, (1, 2, ..., d), \\
    -1\,\, \text{if}\,\, (\m_1, \m_2, ..., \m_d)\,\, \text{is an odd permutation of}\,\, (1, 2, ..., d),  \\
    0\,\,\text{otherwise}\,,
    \end{cases}
\end{align}
and the Levi-Civita tensor as follows:
\begin{align}\label{levi_tens_symb}
    \epsilon^{\m_1\m_2...\m_d}=\frac{\text{sgn}(g)}{\sqrt{\abs{g}}}\,{\hat\varepsilon}^{\m_1\m_2...\m_d}\,,
\end{align}
where $g$ is the determinant of the metric and $\text{sgn}(g)$ is its signature.

The generalized Kronecker delta function (gKd) is defined as
\begin{align}\label{gKd_as_det}
    \d^{\m_1\m_2...\m_p}_{\n_1\n_2...\n_p} &=\left.\begin{vmatrix}
\d^{\m_1}_{\n_1} & \cdots & \d^{\m_p}_{\n_p} \\
\vdots & \ddots & \vdots \\ 
\d^{\m_p}_{\n_1} & \cdots & \d^{\m_p}_{\n_p} 
\end{vmatrix}
\right.\,,
\end{align}
where $\d^{\m_1}_{\n_1}$ is the Kronecker Delta function. The gKd is related to the Levi-Civita tensor as follows:
\begin{align}    
    \epsilon^{\m_1...\m_p\l_{p+1}...\l_d}\,\epsilon_{\n_1...\n_p\l_{p+1}...\l_d}=\sgn(g)\,(d-p)!\,\d^{\m_1\m_2...\m_p}_{\n_1\n_2...\n_p}. \label{eps_partial_contr}
\end{align}

\subsection{Differential Forms}
\label{sec:forms-appendix}

% In the literature on M5-branes, the opposite convention is used. The differential $p-$form is defined differently, and the interior and exterior product applies on the right. examples of papers explicitly using this convention (the list will be added later if other refs are found): hep-th/9703127, hep-th/0409107 (the reason for using non-standard convention is not simply historical, but it is actually the convention suitable to work with superspace calculations. See last paragraph of page 5 of this paper), 1406.5185 (see section 2 of this paper)

% standard superspace conventions. Since the fermionic property of some components and
% differentials give signs depending on ordering, this is a convenient way of handling these
% with a minimum of extra signs.

In the following, the expressions shown inside the big round brackets are the expressions used in index notation. \\ \\
\textbf{$\bf p$-form}
\begin{align}\label{p-form}
    &A^{(p)}=\frac{1}{p!}\,\left(A_{[\m_1\m_2...\m_p]}\right)\,dx^{\m_1}\w dx^{\m_2}\w...dx^{\m_p}.
\end{align}
\textbf{Exterior Product}
\begin{align}
    &\quad\, A^{(p)}\w B^{(q)}\w...\w C^{(r)} \nn\\
    &=\frac{1}{(p+q+...\,r)!}\left(\frac{(p+q+...\,r)!}{p!\,q!\,...\,r!}\,A_{[\m_1...\m_p}B_{\n_1...\n_q}...C_{\l_1...\l_r]}\right)dx^{\m_1}\w ...\,dx^{\m_p}\w dx^{\n_1}\w\qquad\qquad\qquad\nn\\    &\qquad\qquad\qquad\qquad\qquad\qquad\qquad\qquad\qquad\qquad\qquad\qquad\quad\,...\,dx^{\n_q}...\,\w dx^{\l_1}\w...\,dx^{\l_r}.\label{AwedgeB}
\end{align}

\textbf{Interior Product}

For a vector field $v=v^{\m_1}\,\frac{\pa}{\,\pa x^{\m_1}}\,$,
\begin{align}\label{int_prod}
    &\i_v(A^{(p)})=\frac{1}{(p-1)!}\,\left(v^{\m_1}A_{[\m_1\m_2...\m_p]}\right)\,dx^{\m_2}\w dx^{\m_3}...\w dx^{\m_p}\,.
\end{align}
\textbf{Hodge Dual}
\begin{align}
    &\quad{*}(A^{(p)})=(* A)^{(d-p)}=\frac{1}{(d-p)!}\left(\frac{1}{p!}\,\epsilon_{\m_1 \m_2...\m_d}\,A^{\m_1\m_2...\m_p}\right)dx^{\m_{(p+1)}}\w...\,d x^{\m_d}\,. \label{hodged} \\
    &\quad{*}(A^{(p)}\w B^{(q)}\w...\w C^{(r)}) \nn\\
    &=\frac{1}{n!}\left(\frac{1}{p!\,q!\,...\,r!}\,\e_{\m_1..\m_p\n_1...\n_q...\l_1...\l_r\sigma_1...\sigma_n} A^{\m_1...\m_p}B^{\n_1...\n_q}...\,C^{\l_1...\l_r}\right) dx^{\sigma_1}\w...\,dx^{\sigma_n}\,.
\end{align}
where $n=d-p-q-...\,r$.

\textbf{Exterior Product of $p$-form with Hodge Dual of $p$-form}
\begin{align}
    &\quad\, A^{(p)}\w* (B^{(p)})=A^{(p)}\w(* B)^{(d-p)} \nonumber \\
    &=\frac{1}{d!}\left(\frac{d!}{(p!)^2\,(d-p)!}\,B^{\n_1...\n_p}\,\epsilon_{\n_1...\n_p[\n_{(p+1)}...\n_d}A_{\m_1...\m_p]}\right)dx^{\m_1}\w...dx^{\m_p}\w dx^{\n_{(p+1)}}\w... dx^{\n_d}\, \label{1stline}\\
    &=\frac{1}{d!}\left(\frac{1}{p!}\,\epsilon_{\m_1...\m_p\n_{(p+1)}...\n_d}\,A_{\n_1...\n_p}\,B^{\n_1...\n_p}\right)dx^{\m_1}\w...dx^{\m_p}\w dx^{\n_{(p+1)}}\w... dx^{\n_d}. \label{2ndline}
\end{align}

\textbf{Exterior Derivative}
\begin{align}
    &d(A^{(p)})=\frac{1}{(p+1)!}\left((p+1)\,\pa_{[\m_1}A_{\m_2...\m_{(p+1)}]}\right)\,dx^{\m_1}\w dx^{\m_2}\w...dx^{\m_{(p+1)}}.
\end{align}

\textbf{Integral}

The volume element is
\begin{align}\label{epsilon-d}
    \epsilon^{(d)}=\frac{1}{d!}\,\epsilon_{\m_1 \m_2 ...\m_d}\,dx^{\m_1} \w  dx^{\m_2}\w...\w dx^{\m_d}=\sqrt{\abs{g}}\,dx^0 \w dx^1\w ...\w dx^{(d-1)}\equiv(\sqrt{\abs{g}}\,d^dx).
\end{align}
On a $d$-dimensional manifold, the integral of a $d$-form is as follows:
\begin{align}
    \int C^{(d)}\equiv\int \epsilon^{(d)}\,(* C)^{(0)}\equiv\int d^dx\, \frac{\sqrt{\abs{g}}}{\,d!}\,\epsilon_{\m_1 \m_2 ...\m_d}C^{\m_1 \m_2 ...\m_d}.
\end{align}

\subsection{Exterior Algebra Identities}\label{ext_algeb}

We take a 0-form $a$, two normalised vector fields $u$ and $v$, and two $p$-forms $A$ and $B$. The number of dimensions is $d$. For a vector field $v=v^\m\,\pa_\m\,$, the corresponding $1$-form $v_\m\,dx^\m\,$ is denoted by $\,v^\flat$\footnote{referred to as "$v$-flat" adopting a musical isomorphism}.
\begin{align}
    &{**}A=\text{sgn}(g)(-1)^{p(d-p)}A\,, \label{starstarid}\\
    &A\w{*}B= B\w {*}A\,, \label{AstarC}\\
    &\i_v{*A} =*(A\w v^\flat)\,, \label{ivstarid}\\
    &{*}(\i_v A)=(-1)^{(p-1)}\, v^\flat \w{*}A =(-1)^{(d-1)}\,(*A)\w v^\flat\,, \label{starivid} \\
    &\i_u(v^\flat\w A)+v^\flat\w \i_uA=(u^\m\,v_\m)\,A\,, \label{projrej}\\
    &Q^{(d)}=\frac{v^\flat\w \i_uQ}{(u^\m\,v_\m)}\,, \label{top-form}\\
    &\i_v(u^\flat\w{*A})=*(v^\flat\w\i_uA)\,, \label{starviu}\\
    &v^\flat\w\i_u{*A}=*\i_v(u^\flat\w A)\,. \label{starivu}
\end{align}
% \begin{align}
%     {**A}&=\text{sgn}(g)(-1)^{p(d-p)}A\,, \label{starstarid}\\
%     A\w {*C}&= C\w {*A} \label{AstarC}\\
%     \i_v{*A} &=*(A\w v^\flat) \label{ivstarid}\\
%     *(\i_v A)&=(-1)^{(p-1)}\, v^\flat \w*A =(-1)^{(d-1)}\,(*A)\w v^\flat \label{starivid} \\
%     \i_u(v^\flat\w A)+v^\flat\w \i_uA&=(u^\m\,v_\m)\,A \label{projrej} \quad \text{(projection rejection decomposition)} \\
%     M^{(d)}&=\frac{v^\flat\w \i_uM}{(u.v)} \label{top-form}\\
%     \i_v(u^\flat\w{*A})&=*(v^\flat\w\i_uA) \label{starviu}\\
%     v^\flat\w\i_u{*A}&=*\i_v(u^\flat\w A) \label{starivu}
% \end{align}

\section{Recasting Certain 5-form Expressions}\label{app:attemptEa}

Let $v$ be a normalised vector field such that $\i_v v^\flat =v^\m\,v_\m = 1.$
Let $P$ be a $2$-form such that $\i_v P = 0.$
So $P = \i_v(v^\flat\w P).$
Define $u^\flat\equiv *(v^\flat\w P\w P).$
Note that $\i_v u^\flat = 0.$ To show this, consider
\begin{align}
\i_v u^\flat=\i_v {*}(v^\flat\w P\w P)=-\,{*}(v^\flat\w v^\flat\w P\w P)= 0\,.
\end{align}
Since in six dimensions antisymmetrisation over seven indices is trivially zero, we have 
\begin{align}
    7v_{[\m_0} P_{\m_1\m_2}P_{\m_3\m_4} P_{\m_5\m_6]}
    = v_{\m_0} P_{[\m_1\m_2}P_{\m_3\m_4} P_{\m_5\m_6]}
    + 6\,v_{[\m_1} P_{\m_2\m_3}P_{\m_4\m_5} P_{\m_6]\m_0}=0.
\end{align}
Multiplying by $\frac18\,\epsilon^{\m_1\m_2\m_3\m_4\m_5\m_6}\,d x^{\m_0}$, we get
\begin{align}
 &( v_{\m_0} d x^{\m_0}) \lb \tfrac18\,\epsilon^{\m_1\m_2\m_3\m_4\m_5\m_6}
 P_{\m_1\m_2}P_{\m_3\m_4} P_{\m_5\m_6} \rb
 + 3 \lb \tfrac14\, \epsilon^{\m_1\m_2\m_3\m_4\m_5\m_6} v_{\m_1} P_{\m_2\m_3}P_{\m_4\m_5} \rb\! P_{\m_6\m_0} d x^\m_0=0 \nn\\
 \Rightarrow \,\, & {*(P\w P \w P)}\, v^\flat
 + 3\, u^{\m_6} P_{\m_6\m_0} d x^{\m_0}=0 \nn\\
\Rightarrow \,\,&{*(P \w P \w P)}\, v^\flat
    + 3\, \i_u P=0.
\end{align}
% = v^\flat \w\i_v[\i_v(v^\flat\w P)\w\i_v(v^\flat\w P)\w\i_v(v^\flat\w P)]
Furthermore, using \eqref{projrej} and the fact that $P\w P \w P$ is a full-rank form, we have  $P\w P \w P = v^\flat \w \i_v (P\w P \w P)
= 3\,v^\flat \w (\i_v P) \w P \w P = 0$. It implies that 
\begin{align}
    \i_u P = 0.
\end{align}
Then it follows that 
\begin{align}
    {*}[P\w v^\flat \w {*}\{P\w{*}(P\w P)\}] 
    &= -\,{*}[P\w {*}\{\i_v\big(P\w{*}(P\w P)\big)\}] 
    \quad (\text{using \eqref{starivid}})\nn\\
    &= -\,{*}[P\w {*}\{P\w \i_v{*}(P\w P)\}] \nn\\
    &= -\,{*}\{P\w {*}(P\w u^\flat)\} \nn\\
    &= -\,{*}(P\w \i_u{*P})
    \quad (\text{using \eqref{ivstarid}}) \nn\\
    &= -\,{*}\i_u(P\w {*P}) \nn\\
    &= {*}(P\w {*P})\, u^\flat
    \quad (\text{using \eqref{starivid}}).
\end{align}
When $P=\i_v{*H}$ the above identity gives
\begin{align}
    *[v^\flat\w\i_v {*H}\w{*}\{\i_v {* H}\w*(\i_v{*H}\w\i_v{*H})\}]
=[{*}\{\i_v {* H}\w*(\i_v{*H})\}]^{(0)}\,{*}(v^\flat\w\i_v{*H}\w\i_v {*H}),
\end{align}
where the superscript $^{(0)}$ reminds us that the term is a $0$-form/scalar. Taking Hodge dual of both sides, we get
\begin{align}\label{iden_Ea1}
    v^\flat\w\i_v {*H}\w{*}\{\i_v {* H}\w*(\i_v{*H}\w\i_v{*H})\}
=[{*}\{\i_v {* H}\w{*}(\i_v{*H})\}]^{(0)}\,v^\flat\w\i_v{*H}\w\i_v {*H}.
\end{align}
Next we have 
\begin{align}
    &\quad\,
    {*}[v^\flat\w{*}\{P\w{*}(P\w P)\}\w{*}\{P\w{*}(P\w P)\}] \nn\\
    &= -\,{*}[{*}\{P\w\i_v{*(P\w P)}\}\w{*}\{P\w{*}(P\w P)\}]
    \quad (\text{using \eqref{starivid}}) \nn\\
    &= -\,{*}[{*}(P\w u^\flat)\w{*}\{P\w{*}(P\w P)\}] \nn\\
    &= -\,{*}[\i_u{*P}\w{*}\{P\w{*}(P\w P)\}]
    \quad (\text{using \eqref{ivstarid}}) \nn\\
    &= -\,{*}\i_u[{*P}\w{*}\{P\w{*}(P\w P)\}]
    + {*}[{*P}\w \i_u{{*}\{P\w{*}(P\w P)\}}] \nn\\
    &= {*}[{*P}\w{*}\{P\w{*}(P\w P)\}]\, u^\flat
    + {*}[{*P}\w {{*}\{P\w u^\flat\w{*}(P\w P)\}}]
    \quad (\text{using \eqref{starivid} and \eqref{ivstarid}}) \nn\\
    &= {*}[{*P}\w{*}\{P\w{*}(P\w P)\}]\, u^\flat
    - {*}[{*P}\w {{*}\{P\w {*}(2\,\i_uP\w P))}\}]
    \quad (\text{using \eqref{starivid}}) \nn\\
    &= {*}[{*P}\w{*}\{P\w{*}(P\w P)\}]\, u^\flat \nn\\
    &= {*}\{P\w{*}(P\w P)\w{*}({*P})\}\, u^\flat
    \quad (\text{using \eqref{AstarC}}) \nn\\
    &= -\,{*}\{P\w P \w {*}(P\w P)\}\, u^\flat.
\end{align}
When $P=\i_v{*H}$ the above identity gives
\begin{align}
    &\quad{*}[v^\flat\w{*}\{\i_v{* H}\w{*}(\i_v{*H}\w\i_v{*H})\}\w{*}\{\i_v {* H}\w{*}(\i_v{*H}\w\i_v{*H})\}] \nn\\
    &=-\,[{*}\{\i_v{*H}\w\i_v{*H}\w*(\i_v{*H}\w\i_v{*H})\}]^{(0)}\,{*}(v^\flat\w\i_v{*H}\w\i_v{*H}).
\end{align}
The superscript $^{(0)}$ reminds us that the expression inside its bracket is a $0$-form. Taking the Hodge dual of both the sides, we get
\begin{align}\label{iden_Ea2}
    &\,v^\flat\w{*}\{\i_v{* H}\w{*}(\i_v{*H}\w\i_v{*H})\}\w{*}\{\i_v {* H}\w{*}(\i_v{*H}\w\i_v{*H})\} \nn\\
    =&-[{*}\{\i_v{*H}\w\i_v{*H}\w*(\i_v{*H}\w\i_v{*H})\}]^{(0)}\,v^\flat\w\i_v{*H}\w\i_v{*H}.
\end{align}

\section{Recasting a Determinant}\label{determinant}

Leibniz formula for determinants tells us that the determinant of an $n\times n$ matrix $M$ whose elements are denoted by $M_{\m\n}$, is given by 
\begin{align}
    \det(M)&=\frac{1}{n!}\,\hat\varepsilon^{i_1i_2...i_n}\,\hat\varepsilon^{j_1j_2...j_n}\,M_{\m_1\n_1}M_{\m_2\n_2}...\,M_{\m_n\n_n},
\end{align}
where $\hat\varepsilon_{i_1i_2...i_n}$ is the Levi-Civita symbol with $\hat\varepsilon^{12...n}=1$.

So in $6D$-spacetime where $\,g_{\m\n}\,$ is the metric and $\,T_{\m\n}\,$ is a rank-$2$ tensor with indices $\,\mu,\n=0,1,...5\,$, the determinant of $\,(g_{\m\n}+T_{\m\n})\,$ can be given by
\begin{align}\label{det_gT}
    \det(g_{\m\n}+T_{\m\n})=\frac{1}{6!}\,\hat\varepsilon^{\mu_0\ldots\mu_5}\,
  \hat\varepsilon^{\n_0\ldots\n_5}\,(g_{\m_0 \n_0}+T_{\m_0\n_0})
  (g_{\m_1 \n_1}+T_{\m_1\n_1})\ldots (g_{\m_5 \n_5}+T_{\m_5\n_5}).
\end{align}
If $\,T_{\m\n}\,$ is antisymmetric then the above equation can also be written as
\begin{align}
    \det(g_{\m\n}+T_{\m\n})=\frac{1}{6!}\,\hat\varepsilon^{\mu_0\ldots\mu_5}\,
  \hat\varepsilon^{\n_0\ldots\n_5}\,(g_{\m_0 \n_0}-T_{\n_0\m_0})
  (g_{\m_1 \n_1}-T_{\n_1\m_1})\ldots (g_{\m_5 \n_5}-T_{\n_5\m_5}).
\end{align}
As $\,g_{\m\n}\,$ is a symmetric tensor, we can exchange its indices without getting any change in its sign. Performing this exchange on the right hand side (RHS) of the above equation we get
\begin{align}\label{det_even}
    \det(g_{\m\n}+T_{\m\n})=\,\,&\frac{1}{6!}\,\hat\varepsilon^{\mu_0\ldots\mu_5}\,
  \hat\varepsilon^{\n_0\ldots\n_5}\,(g_{\n_0 \m_0}-T_{\n_0\m_0})
  (g_{\n_1 \m_1}-T_{\n_1\m_1})\ldots (g_{\n_5 \m_5}-T_{\n_5\m_5})\,, \nn\\
  =\,\,&\det(g_{\m\n}-T_{\m\n}).
\end{align}
As $\,\det(g_{\m\n}+T_{\m\n})\,$ is even under $T_{\m\n}\rightarrow -T_{\m\n}$, the expansion of $\,\det(g_{\m\n}+T_{\m\n})\,$ consists of only the even powers of $T_{\m\n}$. We continue to work with $\,T_{\m\n}\,$ as an anti-symmetric tensor. Using binomial expansion, \eqref{det_gT} can be expressed as
\begin{align}
    &\quad\,\det(T_{\m\n}+g_{\m\n})\nn\\
    &=\sum_{n=0}^6 \frac{6!}{n!\,(6-n)!}
    \frac{1}{6!}\,\hat\varepsilon^{\mu_0\ldots\mu_5}\,
  \hat\varepsilon^{\n_0\ldots\n_5}\,
    (T_{\m_0\n_0} \ldots T_{\m_{(n-1)}\n_{(n-1)}})
    ( g_{\m_n\n_n} \ldots g_{\m_5\n_5}) \nn\\
    &=|g|\sum_{n=0}^6 \frac{6!}{n!\,(6-n)!}
    \frac{1}{6!}\,\epsilon^{\mu_0\ldots\mu_5}\,
  \epsilon^{\n_0\ldots\n_5}\,
    (T_{\m_0\n_0} \ldots T_{\m_{(n-1)}\n_{(n-1)}})
    ( g_{\m_n\n_n} \ldots g_{\m_5\n_5}) \quad \text{(using }\eqref{levi_tens_symb}\text{)} \nn\\
    &=|g|\sum_{n=0}^6 \frac{1}{n!\,(6-n)!}\,\epsilon^{\mu_0\ldots\mu_5}\,
  \epsilon_{\n_0\ldots\n_{(n-1)}\m_n\ldots\m_{5}}\,
    T_{\m_0}{}^{\n_0} \ldots T_{\m_{(n-1)}}{}^{\n_{(n-1)}}  \nn\\
    &=|g|\sum_{n=0}^6 \frac{\sgn(g)}{n!}\,\d^{\m_0\ldots\m_{(n-1)}}_{\n_0\ldots\n_{(n-1)}}\,T_{\m_0}{}^{\n_0} \ldots T_{\m_{(n-1)}}{}^{\n_{(n-1)}} \quad \text{(using }\eqref{eps_partial_contr}\text{)} \nn\\
    &= g\,[ 1 +  \tfrac{1}{2!}\,
    \d^{\m_0\m_1}_{\n_0\n_1}
    T_{\m_0}{}^{\n_0} T_{\m_1}{}^{\n_1}
    + \tfrac{1}{4!}\,\d^{\m_0\m_1\m_2\m_3}_{\n_0\n_1\n_2\n_3}
    T_{\m_0}{}^{\n_0} T_{\m_1}{}^{\n_1}T_{\m_2}{}^{\n_2}T_{\m_3}{}^{\n_3} \nn\\
    &\qquad\qquad\,\, + \tfrac{1}{6!}\,\d^{\m_0\m_1\m_2\m_3\m_4\m_5}_{\n_0\n_1\n_2\n_3\n_4\n_5}
    T_{\m_0}{}^{\n_0} T_{\m_1}{}^{\n_1}T_{\m_2}{}^{\n_2}T_{\m_3}{}^{\n_3}T_{\m_4}{}^{\n_4}T_{\m_5}{}^{\n_5}]. \label{det_gKd_exp}
\end{align}
In the last line above we have used the fact that $|g|=-g$ and $\,\sgn(g)=-1$ for our Minkowski metric $\,g_{\m\n}$ with mostly-plus signature. We simplify each term on the RHS separately.
\begin{align}
   &\quad\,\tfrac{1}{2!}\,\delta^{\m_0\m_1}_{\n_0\n_1}
    T_{\m_0}{}^{\n_0} T_{\m_1}{}^{\n_1} \nn\\
    &= \tfrac12\,( \delta^{\m_0}_{\n_0} \delta^{\m_1}_{\n_1} -
\delta^{\m_0}_{\n_1} \delta^{\m_1}_{\n_0} )\,
  T_{\m_0}{}^{\n_0} T_{\m_1}{}^{\n_1}  \nn\\
  &= \tfrac12\,T_{\m_0\m_1} T^{\m_0\m_1} \nn\\
  &= \tfrac14\,\d^{\m_0\m_1}_{\n_0\n_1}T_{\m_0\m_1} T^{\n_0\n_1} \nn\\
  &=\sgn(g)\,\tfrac{1}{96}\,\epsilon^{\m_0\m_1\m_2\m_3\m_4\m_5}\{T_{\m_0\m_1}(\epsilon_{\n_0\n_1\m_2\m_3\m_4\m_5} T^{\n_0\n_1})\} \quad \text{(using }\eqref{eps_partial_contr}\text{)} \nn\\
  &=-\,\tfrac{1}{96}\,\epsilon^{\m_0\m_1\m_2\m_3\m_4\m_5}\big\{\tfrac{(2!)^2\,4!}{6!}\,(T\w{*T})\big\} \quad \text{(using }\eqref{1stline}\text{)}\nn\\
  &=-\,\tfrac{1}{96}\,\tfrac{6!\,(2!)^2\,4!}{6!}\,{*}(T\w{*T}) \quad \text{(using }\eqref{2ndline}\text{)} \nn\\
    &=-\, {*}(T\w {* T}) \quad \text{(in differential form notation)}.\label{det_term1}
\end{align}
%    &=\tfrac18\,\d^{\m_0\m_1\m_2\m_3\m_4\m_5}_{\n_0\n_1\m_2\m_3\m_4\m_5}T_{\m_0\m_1} T^{\n_0\n_1} \nn\\
From \eqref{gKd_as_det} we know that
\begin{align}\label{4d_gKd_expan}
 &\,\delta^{\m_0\m_1\m_2\m_3}_{\n_0\n_1\n_2\n_3}\nn\\
  =\,\,& \delta^{\m_0}_{\n_0} \delta^{\m_1}_{\n_1} \delta^{\m_2}_{\n_2} \delta^{\m_3}_{\n_3}
    - \delta^{\m_0}_{\n_0} \delta^{\m_1}_{\n_1} \delta^{\m_2}_{\n_3} \delta^{\m_3}_{\n_2}
    - \delta^{\m_0}_{\n_0} \delta^{\m_1}_{\n_2} \delta^{\m_2}_{\n_1} \delta^{\m_3}_{\n_3}
    + \delta^{\m_0}_{\n_0} \delta^{\m_1}_{\n_2} \delta^{\m_2}_{\n_3} \delta^{\m_3}_{\n_1}
    + \delta^{\m_0}_{\n_0} \delta^{\m_1}_{\n_3} \delta^{\m_2}_{\n_1} \delta^{\m_3}_{\n_2}
    - \delta^{\m_0}_{\n_0} \delta^{\m_1}_{\n_3} \delta^{\m_2}_{\n_2} \delta^{\m_3}_{\n_1}
  \nn\\
  &- \delta^{\m_0}_{\n_1} \delta^{\m_1}_{\n_0} \delta^{\m_2}_{\n_2} \delta^{\m_3}_{\n_3}
    + \delta^{\m_0}_{\n_1} \delta^{\m_1}_{\n_0} \delta^{\m_2}_{\n_3} \delta^{\m_3}_{\n_2}
    + \delta^{\m_0}_{\n_1} \delta^{\m_1}_{\n_2} \delta^{\m_2}_{\n_0} \delta^{\m_3}_{\n_3}
    - \delta^{\m_0}_{\n_1} \delta^{\m_1}_{\n_2} \delta^{\m_2}_{\n_3} \delta^{\m_3}_{\n_0}
    - \delta^{\m_0}_{\n_1} \delta^{\m_1}_{\n_3} \delta^{\m_2}_{\n_0} \delta^{\m_3}_{\n_2}
    + \delta^{\m_0}_{\n_1} \delta^{\m_1}_{\n_3} \delta^{\m_2}_{\n_2} \delta^{\m_3}_{\n_0}
    \nn\\
  &+ \delta^{\m_0}_{\n_2} \delta^{\m_1}_{\n_0} \delta^{\m_2}_{\n_1} \delta^{\m_3}_{\n_3}
    - \delta^{\m_0}_{\n_2} \delta^{\m_1}_{\n_0} \delta^{\m_2}_{\n_3} \delta^{\m_3}_{\n_1}
    - \delta^{\m_0}_{\n_2} \delta^{\m_1}_{\n_1} \delta^{\m_2}_{\n_0} \delta^{\m_3}_{\n_3}
    + \delta^{\m_0}_{\n_2} \delta^{\m_1}_{\n_1} \delta^{\m_2}_{\n_3} \delta^{\m_3}_{\n_0}
    + \delta^{\m_0}_{\n_2} \delta^{\m_1}_{\n_3} \delta^{\m_2}_{\n_0} \delta^{\m_3}_{\n_1}
    - \delta^{\m_0}_{\n_2} \delta^{\m_1}_{\n_3} \delta^{\m_2}_{\n_1} \delta^{\m_3}_{\n_0}
    \nn\\
  &- \delta^{\m_0}_{\n_3} \delta^{\m_1}_{\n_0} \delta^{\m_2}_{\n_1} \delta^{\m_3}_{\n_2}
    + \delta^{\m_0}_{\n_3} \delta^{\m_1}_{\n_0} \delta^{\m_2}_{\n_2} \delta^{\m_3}_{\n_1}
    + \delta^{\m_0}_{\n_3} \delta^{\m_1}_{\n_1} \delta^{\m_2}_{\n_0} \delta^{\m_3}_{\n_2}
    - \delta^{\m_0}_{\n_3} \delta^{\m_1}_{\n_1} \delta^{\m_2}_{\n_2} \delta^{\m_3}_{\n_0}
    - \delta^{\m_0}_{\n_3} \delta^{\m_1}_{\n_2} \delta^{\m_2}_{\n_0} \delta^{\m_3}_{\n_1}
    + \delta^{\m_0}_{\n_3} \delta^{\m_1}_{\n_2} \delta^{\m_2}_{\n_1} \delta^{\m_3}_{\n_0}.
\end{align}
Kronecker-delta $\d_\m^\n$ is symmetric in $\m,\n$ whereas $T_\m{}^\n$ is anti-symmetric in $\m,\n$. Therefore when the above twenty-four terms are contracted with $T_{\m_0}{}^{\n_0} T_{\m_1}{}^{\n_1}T_{\m_2}{}^{\n_2}T_{\m_3}{}^{\n_3}$, fifteen of them give $0$. We are left with
\begin{align}
&\quad\,\,\tfrac{1}{4!}\,\delta^{\m_0\m_1\m_2\m_3}_{\n_0\n_1\n_2\n_3}
  T_{\m_0}{}^{\n_0} T_{\m_1}{}^{\n_1} T_{\m_2}{}^{\n_2} T_{\m_3}{}^{\n_3} \nn\\
  &=\tfrac{1}{24} 
    (\delta^{\m_0}_{\n_1} \delta^{\m_1}_{\n_0} \delta^{\m_2}_{\n_3} \delta^{\m_3}_{\n_2}
    - \delta^{\m_0}_{\n_1} \delta^{\m_1}_{\n_2} \delta^{\m_2}_{\n_3} \delta^{\m_3}_{\n_0}
    - \delta^{\m_0}_{\n_1} \delta^{\m_1}_{\n_3} \delta^{\m_2}_{\n_0} \delta^{\m_3}_{\n_2}
    - \delta^{\m_0}_{\n_2} \delta^{\m_1}_{\n_0} \delta^{\m_2}_{\n_3} \delta^{\m_3}_{\n_1}
    + \delta^{\m_0}_{\n_2} \delta^{\m_1}_{\n_3} \delta^{\m_2}_{\n_0} \delta^{\m_3}_{\n_1}
    - \delta^{\m_0}_{\n_2} \delta^{\m_1}_{\n_3} \delta^{\m_2}_{\n_1} \delta^{\m_3}_{\n_0}
    \nn\\
  & \qquad\,\,\,- \delta^{\m_0}_{\n_3} \delta^{\m_1}_{\n_0} \delta^{\m_2}_{\n_1} \delta^{\m_3}_{\n_2}
    - \delta^{\m_0}_{\n_3} \delta^{\m_1}_{\n_2} \delta^{\m_2}_{\n_0} \delta^{\m_3}_{\n_1}
    + \delta^{\m_0}_{\n_3} \delta^{\m_1}_{\n_2} \delta^{\m_2}_{\n_1} \delta^{\m_3}_{\n_0})\, T_{\m_0}{}^{\n_0} T_{\m_1}{}^{\n_1} T_{\m_2}{}^{\n_2} T_{\m_3}{}^{\n_3} \nn\\
   &= \tfrac{1}{24}(3\,T_{{\m_0}{\m_1}} T^{{\m_0}{\m_1}}T_{{\n_0}{\n_1}} T^{{\n_0}{\n_1}} - 6\,T_{\m_0\m_1} T^{\m_1\n_0} T_{\n_0\n_1}T^{\n_1\m_0}) \nn\\
   &= \tfrac{1}{8}\,T_{{\m_0}{\m_1}}T_{\n_0\n_1}( T^{{\m_0}{\m_1}} T^{{\n_0}{\n_1}} -2\,T^{\m_1\n_0}T^{\n_1\m_0}) \nn\\
   &= \tfrac{3}{8}\, T_{\m_0\m_1} T_{\n_0\n_1} T^{[\m_0\m_1} T^{\n_0\n_1]} \nn\\
    &= \tfrac{1}{64}\,T_{\m_0\m_1}T_{\n_0\n_1}\,\d^{\m_0\m_1 \n_0\n_1}_{\l_0\l_1\l_2\l_3}\,T^{\l_0\l_1} T^{\l_2\l_3} \nn\\
    &=\sgn(g)\,\tfrac{1}{128}\,\epsilon^{\m_0\m_1 \n_0\n_1\l_4\l_5}\{T_{\m_0\m_1}T_{\n_0\n_1}(\epsilon_{\l_0\l_1\l_2\l_3\l_4\l_5}T^{\l_0\l_1} T^{\l_2\l_3})\} \quad \text{(using }\eqref{eps_partial_contr}\text{)} \nn\\
    &=-\,\tfrac{1}{128}\,\epsilon^{\m_0\m_1 \n_0\n_1\l_4\l_5}\big[\tfrac{(4!)^2\,2!}{6!}\,\big\{\tfrac{(2!)^2}{(2+2)!}\big\}^2\,\{T\w T \w {*(T\w T)}\}\big] \quad \text{(using }\eqref{AwedgeB}\,\,\text{\&}\,\,\eqref{1stline}\text{)} \nn\\
    &=-\,\tfrac{1}{128}\,\tfrac{6!\,(4!)^2\,2!}{6!}\,\big\{\tfrac{(2!)^2}{(2+2)!}\big\}^2\,\{T\w T \w {*(T\w T)}\} \quad \text{(using }\eqref{2ndline}\text{)} \nn\\
    &=-\,\tfrac14\,{*}\{T\w T \w {*(T\w T)}\} \quad \text{(in differential form notation)}. \label{det_term2}
\end{align}
% \nn\\
%     =\,&k T_{\m_0\m_1}T_{\n_0\n_1}\,\d^{\m_0\m_1 \n_0\n_1\l_4\l_5}_{\l_0\l_1\l_2\l_3\l_4\l_5}\,T^{\l_0\l_1}T^{\l_2\l_3} \,\,\text{(using identity \eqref{red_order})} 
Similarly it can be seen that
\begin{align}
&\tfrac{1}{6!}\,\d^{\m_0\m_1\m_2\m_3\m_4\m_5}_{\n_0\n_1\n_2\n_3\n_4\n_5}
    T_{\m_0}{}^{\n_0} T_{\m_1}{}^{\n_1}T_{\m_2}{}^{\n_2}T_{\m_3}{}^{\n_3}T_{\m_4}{}^{\n_4}T_{\m_5}{}^{\n_5} \nn\\
    =\,\,&\sgn(g)\,\tfrac{1}{48^2}\,\epsilon^{\m_0\m_1 \m_2\m_3\m_4\m_5}\{T_{\m_0\m_1}T_{\m_2\n_3}T_{\m_4\n_5}(\epsilon_{\n_0\n_1\n_2\n_3\n_4\n_5}T^{\n_0\n_1} T^{\n_2\n_3}T^{\n_4\n_5})\} \nn\\
    =\,\,&-\tfrac{1}{36}\,{*}\{T\w T \w T\w{*}(T\w T \w T)\} \quad \text{(in differential form notation)}. \label{det_term3}
\end{align}
Incorporating expressions \eqref{det_term1}, \eqref{det_term2} and \eqref{det_term3} into eq. \eqref{det_gKd_exp}, we get
\begin{align}
    &\det(g_{\m\n}+T_{\m\n})\nn\\
    =\,\,&g\,[1 -{*}(T\w {* T})- \tfrac{1}{4} 
  {*}\{T\w T \w {*(T\w T)}\}
    - \tfrac{1}{36}\,{*}\{T\w T \w T\w{*}(T\w T \w T)\}].
\end{align}
On setting $\,T=-\,i\,\i_v{*H}$, we get
\begin{align}
    \det\,(g-i\,\i_v{*H})=&g\,[1 + {*}\{\i_v{*H}\w {*(\i_v{*H})}\}
    - \tfrac{1}{4}\, 
  {*}\{\i_v{*H}\w \i_v{*H} \w {*}(\i_v{*H}\w \i_v{*H})\} \nn\\
  &\quad+\tfrac{1}{36}\,{*}\{\i_v{*H}\w \i_v{*H} \w \i_v{*H}\w{*}(\i_v{*H}\w \i_v{*H} \w \i_v{*H})\}].
\end{align}
Since $\,(\i_v{*H}\w \i_v{*H} \w \i_v{*H})\,$ is a $6$-form in the $6D$ worldvolume of the M5-brane,
by identity \eqref{top-form} it can be written as
\begin{align}
    \i_v{*H}\w \i_v{*H} \w \i_v{*H}=v^\flat\w\i_v(\i_v{*H}\w \i_v{*H} \w \i_v{*H})=0.
\end{align}
Therefore, we find
\begin{align}
    \det\,(g-i\,\i_v{*H})&=g\,[1 + {*}\{\i_v{*H}\w {*(\i_v{*H})}\}
    - \tfrac{1}{4}\, 
  {*}\{\i_v{*H}\w \i_v{*H} \w {*}(\i_v{*H}\w \i_v{*H})\}].
\end{align}

% \addtocontents{toc}{\protect\setcounter{tocdepth}{1}}

\addcontentsline{toc}{section}{References}
\bibliographystyle{JHEP}
\bibliography{references}

\end{document}